\documentclass[twocolumn, times, floatfix]{aastex631}

\usepackage{graphicx}
\usepackage{mathptmx}
\usepackage{orcidlink}
\usepackage{float}
\usepackage[version=3]{mhchem}
\usepackage{url,hyperref,microtype}
\hypersetup{colorlinks, linkcolor=red, citecolor=blue, filecolor=cyan, 
            urlcolor=magenta}

\begin{document}

\title{New Parameter Measurements for the Ultra-Short-Period Planet TOI-1807b}

\author[0000-0002-8486-621X]{Peifeng Peng}
\affiliation{Department of Physics, Durham University, Durham, DH1 3LE, UK}

\author{Hongyi Xiong}
\affiliation{Shenzhen College of International Education, Shenzhen, 518043, People’s Republic of China}

\author{He Li}
\affiliation{School of Physics and Electronic Science, East China Normal University, Shanghai, 200241, People’s Republic of China}

\author{Felix Li}
\affiliation{Vermont Secondary College, Vermont, VIC 3133, Australia}

\author{Tianqi Wang}
\affiliation{The Thacher School, Ojai, CA 93023, USA}

\correspondingauthor{Peifeng Peng}
\email{peifeng.peng@durham.ac.uk}



\begin{abstract}

The ultra-short-period (USP) planets are exoplanets with very short orbital periods ($\emph{P\/} < 1$ day), and TOI-1807b is one such planet recently discovered by the TESS mission where it orbits in the TOI-1807 system that is still little known nowadays. In this paper, we re-analyzed the transit light curves of TOI-1807 using the latest TESS data from Sector 49, combined with previous data from Sector 22 and 23. By running the MCMC simulation through all three sectors, we found that our transit model fits the data from Sector 49 the best, and we deduced that TOI-1807b is a Super-Earth with a mass of $2.27^{+0.49}_{-0.58}\, M_\oplus$, a radius of $1.37^{+0.10}_{-0.09}\, R_\oplus$, a density of $0.875^{+0.264}_{-0.285}\, \rho_\oplus$, and a surface temperature of $1499^{+82}_{-129}\, \mathrm{K}$. We confirmed that TOI-1807b orbits at approximately $0.0135^{+0.0013}_{-0.0022}\, \mathrm{AU}$ with a period of $0.54929^{+0.00012}_{-0.00005}\, \mathrm{days}$, which raises the possibility of the planet being tidally locked due to spin-orbit synchronization. In addition, we renewed an estimate for the conjunction time as $2651.98224^{+0.00112}_{-0.00064}\, \mathrm{BTJD}$. We suggest that TOI-1807b might slowly undergo its orbital decay process, and we further identify that TOI-1807b is in a circular, synchronous orbit and permanently deformed due to tides, leading to $\sim$4\% correction in density. Since TOI-1807 is such a young star with an age of only $300 \pm 80\, \mathrm{Myr}$, we also imply that the radiation emitted from active TOI-1807 could be so intense that it might have destroyed most of the atmosphere over the surface of TOI-1807b.

\end{abstract}



\section{Introduction}
\label{section1}

\begin{figure}[ht!]
   \centering
      \includegraphics{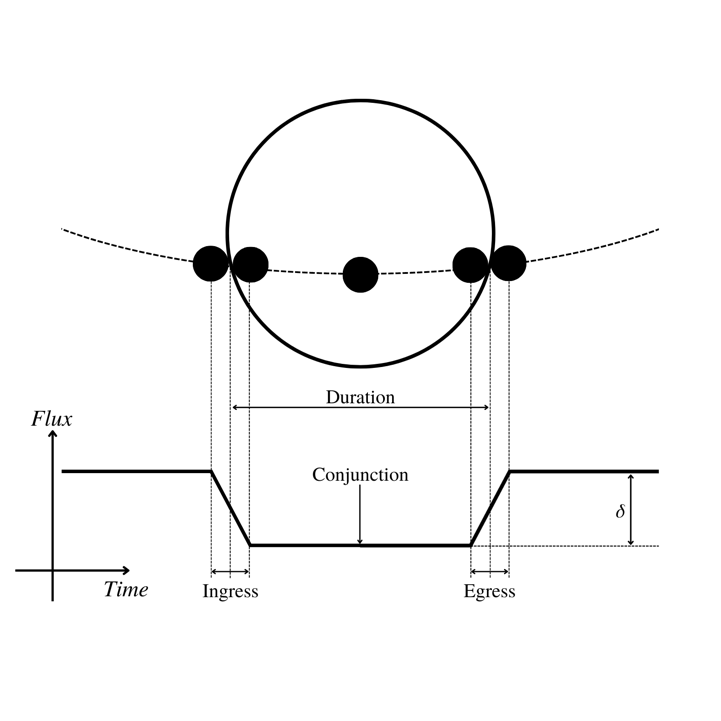}
      \caption{The visible transit of a planet in front of a 
               star recreated from \citet{2010arXiv1001.2010W}. This causes a consequential dip in the flux of the star as the planet covers it partially. Note that the planet’s size has been greatly magnified for visual purposes.
              }
         \label{Fig1}
   \end{figure}

   One of the aims of exoplanet research is to collect data and construct models of other solar systems, which could help scientists come up with possible theories of planet formation from protoplanetary disks. TOI-1807 as a young star with an age of around 300 million years \citep{2022arXiv220603496N} has an orbiting planet TOI-1807b which was recently discovered as the youngest ultra-short-period (USP, $\emph{P\/} < 1$ day) planet, and its existence was originally confirmed by \citet{Hedges_2021}. Take the TOI-1807 system as an example, younger planets with their host younger stars are quite useful for studying formation theory as they have had less time to evolve, therefore retaining more properties of their initial conditions. The space telescopes including Kepler, TESS, and the ongoing James Webb Space Telescope (JWST), have allowed us to gain important information about new exoplanets as well as the detection of thousands. All three telescopes utilize the transit method in order to detect exoplanets, by measuring the flux of a distant star. As a planet orbits in front of the star, there will be a small, but noticeable dip in the brightness emitted by the star. From Fig.~\ref{Fig1} it is possible to determine various planetary parameters such as orbital distance, period, etc.
   
   TESS (short for Transiting Exoplanet Survey Satellite) divides the sky into 26 sectors and studies them over two years, spending approximately 27 days in each sector \citep{2015JATIS...1a4003R}. This is also one of the limitations of TESS as planets with orbital periods longer than 27 days are likely to be excluded from the list of exoplanet discoveries made by TESS as there is not enough time for us to observe a transit of it. A further detailed explanation of the sources of data used in this paper is explained in Section~\ref{section2}. Determining a planet’s transit of its host star isn’t always simple, however, as intrinsic variability of stars as well as USP planets \citep{2018NewAR..83...37W} make precise measurements and transit observation a lot more difficult. The TOI-1807 system fits in both difficulties as the host star TOI-1807 is intrinsically variable and the accompanying planet TOI-1807b is a “lava world” USP. The USP planets are intrinsically rare and have extremely small orbital periods, where they are found to orbit around 0.5\% of Sun-like stars \citep{2014ApJ...787...47S}. This result in the USP planets is one of the earliest surprises in exoplanet astronomy as it challenges our current formation theories. Pre-existing exoplanet formation theories state that planets formed from protoplanetary disks that gradually cool down. Observation of the structures and physical conditions of these protoplanetary disks provides insight into the evolutionary behavior of systems \citep{2020ARA&A..58..483A}.  However, USP planets are an entirely different scenario as the orbital distance from the planet around the star is much too close to allow it to cool down sufficiently, thus requiring modifications to our current theories. This is explained in greater detail in Section~\ref{section5.8}. 

   In this paper, we utilize several methods to update and improve previously calculated parameters of the exoplanet TOI-1807b. These include folding the data to make the transit more obvious as well as the imperative Python packages {\fontfamily{pcr}\selectfont NumPy}, {\fontfamily{pcr}\selectfont Matplotlib}, occultquad from {\fontfamily{pcr}\selectfont mandelagol}, curve fit from {\fontfamily{pcr}\selectfont SciPy}, and {\fontfamily{pcr}\selectfont Lightkurve}. To determine the uncertainty, we used the MCMC (Monte Carlo Markov Chain) method, and the accuracy of the data is analyzed through a reduced $\chi$$^2$ test as well as determining the P-value of the statistic. The MCMC method is split into two parts -- Monte Carlo and Markov Chain. The first part, Monte Carlo, is the process of drawing random samples from a distribution to estimate its properties. The latter part, Markov Chain, is the idea that the random samples used in Monte Carlo are generated via a sequential process based on the previous value \citep{Ravenzwaaij_2018, 2019Ap&SS.364...33H}. In order to improve the fitted equation of the transit, we also flattened the TESS data in single transit intervals, followed by using Curve fit to fit an equation to each of these intervals without using points within the transit. These methods are elaborated in Section~\ref{section3} of the paper. The values we calculated are compared with previously published results in the paper “A precise density measurement of the young ultra-short-period planet TOI-1807b” \citep{2022arXiv220603496N}. We improved upon the calculations in Sector 22 and 23 of TESS which were used by the previous paper, mainly focusing on the parameters of $t_{\mathrm{c}}$ (conjunction time), \emph{P\/} (period), \emph{k\/} (radius of the planet/radius of the star), and \emph{a\/} (radius of the orbit/radius of the star). We also analyzed newer results from Sector 49, which was not covered in previous papers, and compared the data with Sector 22 and 23. This is explained in more detail in Section~\ref{section4} of the paper.
   
   Section~\ref{section5} of the paper discusses multiple calculations regarding the planet’s properties. This includes the calculations of the orbital decay of TOI-1807b (Section~\ref{section5.2}). The period decrease can be explained by tidal orbital decay \citep{2017ApJ...842...40L}. Tidal orbital decay is described as the process when the host star’s gravitational forces raise tides on the planet, causing decay in the planet’s orbit, even when the eccentricity has reached zero \citep{2009IAUS..253..217J}. The second reason is apsidal precession \citep{2017ApJ...842...40L} which is sometimes falsely regarded as orbital decay. This phenomenon is the rotation of the axis connecting the perihelion and aphelion of the planet. Meanwhile, we used Newton-Kepler's laws to calculate the theoretical mass, radius, and density of the planet (Section~\ref{section5.1}) demonstrated in Table~\ref{Table2}. Specifically, the data used in these calculations are from Sector 49, as it has not been previously analyzed, which allows us to compare them to previously published results by \citet{2022arXiv220603496N}. These values are calculated from data in Table~\ref{Table1} as well as MCMC parameters. Next, the planet’s surface temperature is calculated with the Stefan-Boltzmann law, which is done through an estimated albedo from comparison with other USP planets, and previous values of the star \citep{Hedges_2021} from the Planetary Systems Table \citep{ps_2022} in the NASA Exoplanet Archive (DOI: \href{https://catcopy.ipac.caltech.edu/dois/doi.php?id=10.26133/NEA12}{10.26133/NEA12}). We investigate the possible spin-orbit alignment in the TOI-1807 system as well as the nodal precession of TOI-1807b (Section~\ref{section5.4}). We further verify that the orbit of TOI-1807b can not only be assumed as circularized but also synchronous (Section~\ref{section5.3} and~\ref{section5.5}). Another phenomenon explained in this paper is the tidal deformation of TOI-1807b from its host star’s gravity (Section~\ref{section5.6}). Because of the close proximity of the planet to its host star, the gravitational attraction between the star and the planet will be much greater according to the inverse square law. This results in an elongation of the planet's axis in the direction of the star, as well as spin-orbit synchronization. This calculation aims to correct the potential error in the previous paper \citep{2022arXiv220603496N}, which assumes a spherical model for the planet. In Section~\ref{section5.7}, we also discuss the GR (General Relativistic) effects on the planet due to its apsidal precession, which was famously done on Mercury by Einstein.


\section{Source of Data}
\label{section2}

   We obtained our data from TESS and looked at the brightness level of each data point to analyze the star and extract the parameters about TOI-1807. The TESS cameras have an exposure time of 2 seconds, and due to limited storage, images taken by TESS are summed up with short and long cadences before being sent back to Earth. The images of preselected stars are summed up for 2 minutes cadence, while the images of all the stars in the field of view are summed up for 30 minutes cadence \citep{2015JATIS...1a4003R}. TOI-1807 is on the list of the preselected stars, so it was observed with both 2 minutes cadence and 30 minutes cadence. In this paper, we worked with 2 minutes cadence because a shorter time interval allows us to better analyze the transit and optimize our results.

\begin{figure}[ht!]
   \centering
   \includegraphics{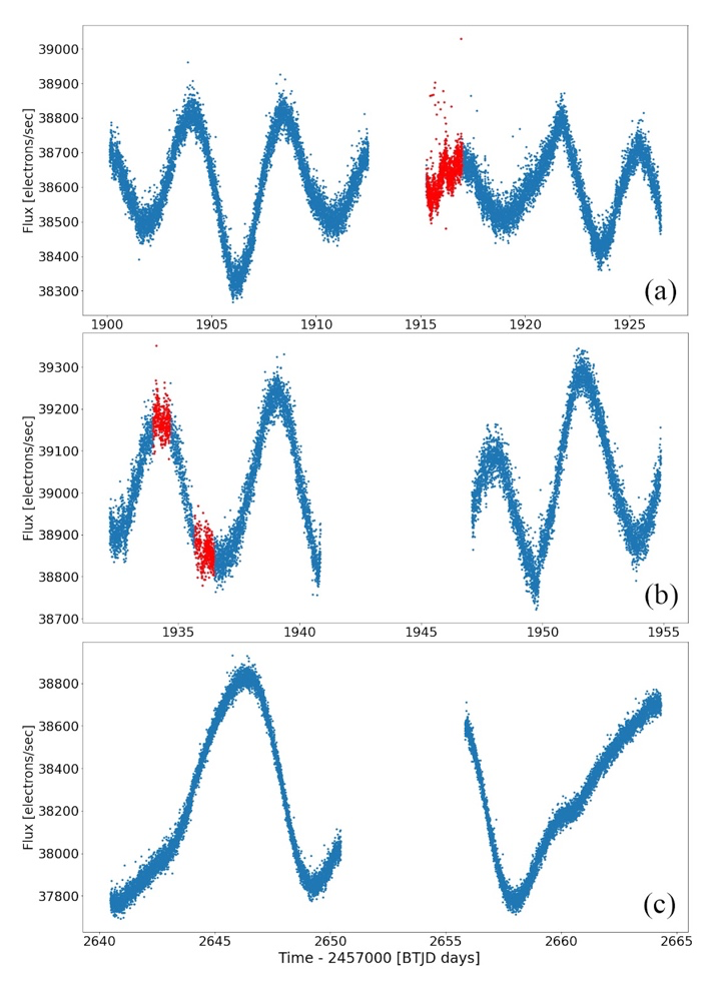}
   \caption{Unprocessed light curve of TOI-1807 from Sector 22, 23, and 49 that we worked with in our research. All the data points are obtained with the 2-minute cadence from TESS and initially processed by the Science Processing and Operation Center (SPOC). Multiple dips in brightness outside of the transit are highlighted in red. Panel (a) shows the light curve from Sector 22, panel (b) shows the light curve from Sector 23, and panel (c) shows the light curve from Sector 49.}
   \label{Fig2}
   \end{figure}

   After TESS images are sent back to Earth, the images are then processed initially by different software to extract the brightness of the selected stars. The data that we used from Sectors 22, 23, and 49 are all produced by the Science Processing and Operation Center (SPOC) and can be found in more detail in MAST: \href{https://archive.stsci.edu/doi/resolve/resolve.html?doi=10.17909/eqx2-n546}{10.17909/eqx2-n546}. We utilized a Python package {\fontfamily{pcr}\selectfont Lightkurve}, which was developed for studying Kepler and TESS data \citep{2018ascl.soft12013L}, to process our data. The package is used in creating a plot of the TOI-1807’s flux over the change of time and other processing, including flattening and folding the light curve. Our research and results are mainly based on the data from Sector 49 to improve on the studies already done for Sector 22 and 23. Sector 49 contains 13275 data points, and the data is observed from 2022 February 26 to 2022 March 26. In addition to the data from Sector 49, we also looked at the data from Sector 22 and Sector 23 in order to compare and validate our results across all sectors. Sector 22 has 16814 data points and spans from 2020 February 19 to 2020 March 17, while Sector 23 has 11072 data points and spans from 2020 March 19 to 2020 April 15. The initial light curve of TOI-1807 from Sectors 22, 23, and 49 are shown in Fig.~\ref{Fig2}.

   While observing the initial light curves from Sector 22 and Sector 23, we found several areas outside of the transits showcasing small dips in the stellar flux, so we initially thought this might be potentially caused by a second transiting exoplanet. The areas of the small dips in brightness are highlighted in red in Fig.~\ref{Fig2}. In testing whether the small dips outside the transits are caused by a second transiting planet, we looked at the pixel file images of these dips, using the “interact” method from the {\fontfamily{pcr}\selectfont Lightkurve} package. This method allows us to access the pixel files of our target star TOI-1807. We detected that many of the pixel files around these dips are empty and do not contain any brightness information (which is very rare for TESS), thus contributing to the brightness fluctuations in these areas. We concluded from the pixel files that these dimming in brightness outside of the transit are not caused by a potential second transiting exoplanet, but by artificial distortion. In addition, these dips in brightness are not periodic and do not follow a certain pattern, further providing evidence that they are unlikely to be a result of another transiting exoplanet.

\begin{figure*}[ht!]
   \centering
   \includegraphics[width=17.3cm]{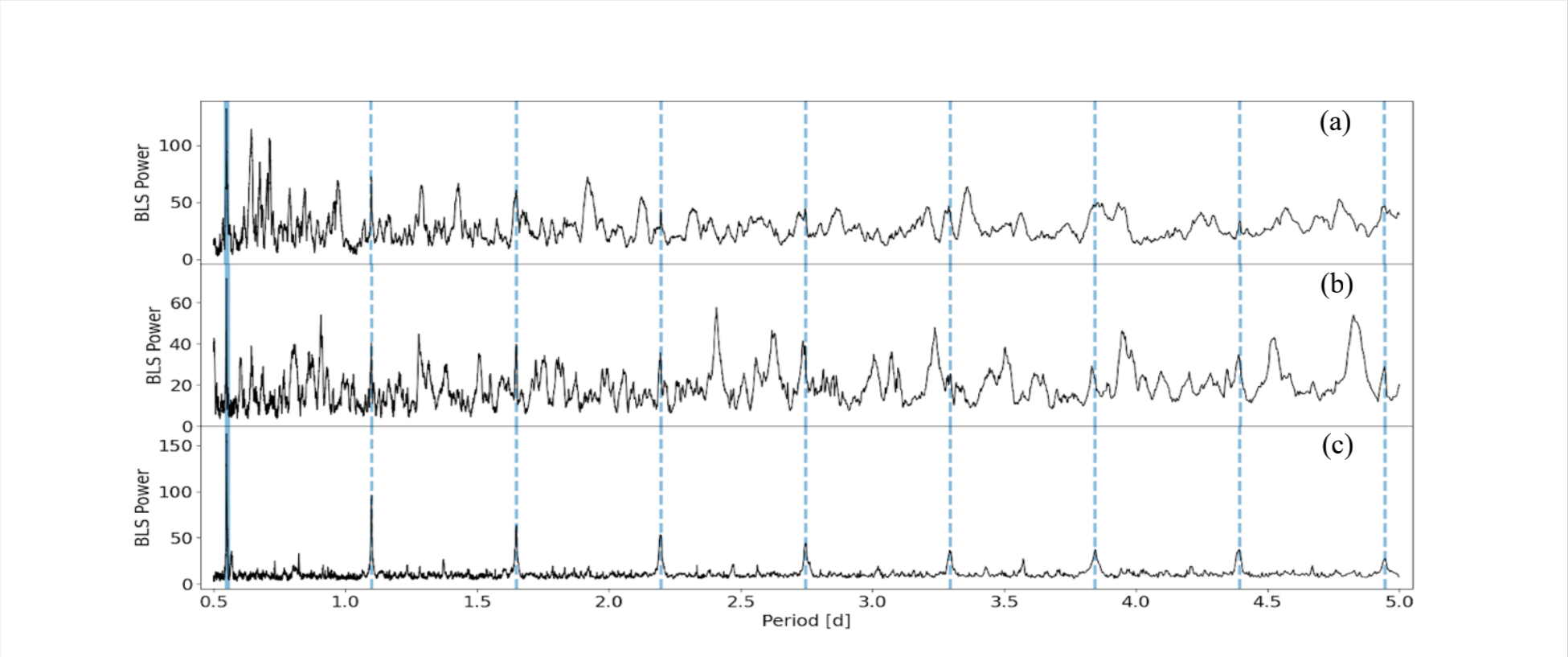}
   \caption{The BLS spectra of the flattened light curves of three sectors to search for periods from 0.5 d to 5.0 d, where panel (a) is for Sector 22, panel (b) is for Sector 23, and panel (c) is for Sector 49. The period corresponding to its highest peak in the periodogram is highlighted with a thick blue line, while other integer harmonics of this period are marked with dashed blue lines. All the thick blue lines and dashed blue lines are aligned well with each other.}
   \label{Fig3}
   \includegraphics[width=17.3cm]{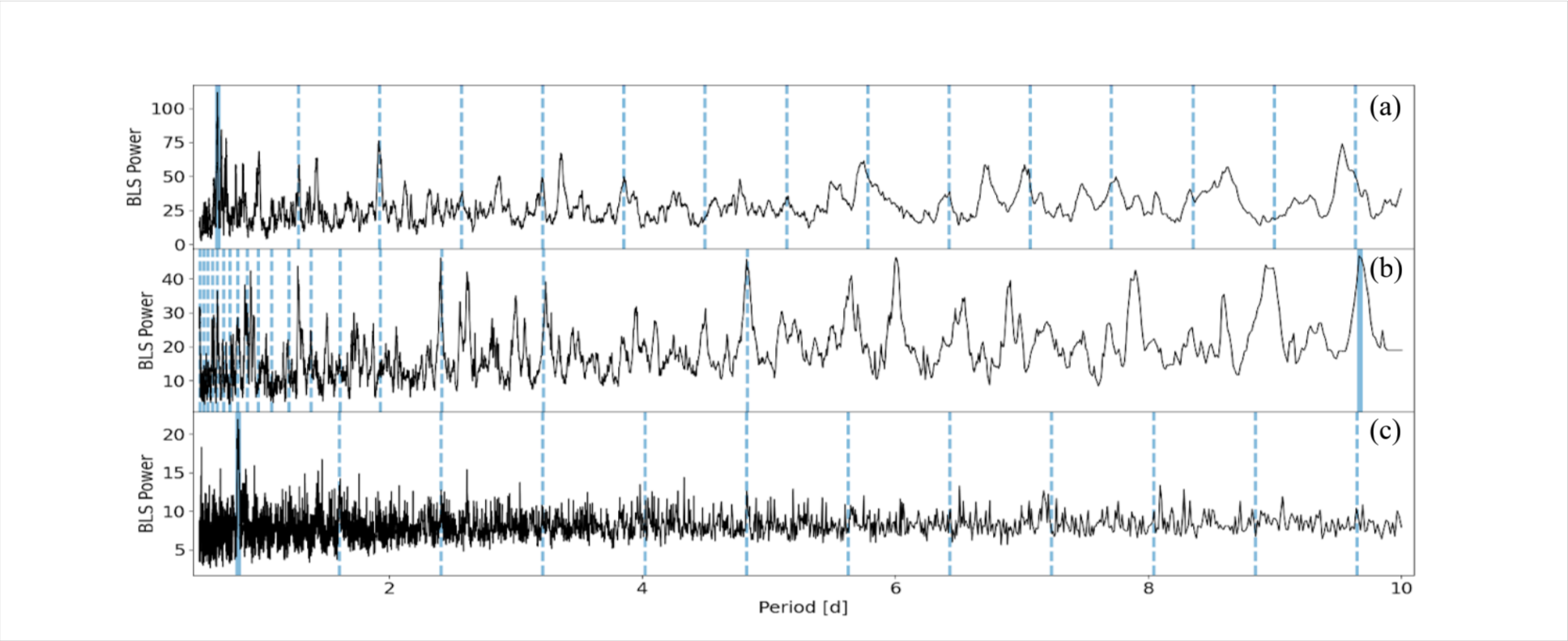}
   \caption{The BLS spectra of the flattened light curves of three sectors to search for periods from 0.5 d to 10.0 d after masking the transit signal of TOI-1807b, where panel (a) is for Sector 22, panel (b) is for Sector 23, and panel (c) is for Sector 49. Neither the thick blue lines nor the dashed blue lines are aligned with each other.}
   \label{Fig4}
   \end{figure*}

   Another more robust method to help us detect any possible transiting exoplanet in the TOI-1807 system is called the Box Least Squares (BLS) analysis, which shows the statistical characteristics of the box-fitting algorithm by modeling each transit as a rectangular dip. This method was first developed by \cite{2002A&A...391..369K}, and it could be used to indicate the transiting exoplanet by identifying its correct period corresponding to the highest peak in the periodogram even at a high noise level. The BLS spectra we obtained by using Python are demonstrated in Fig.~\ref{Fig3}, where one could clearly see three peak periods aligning well with each other at around 0.549 d in three periodograms. To be more specific, with the help of the box-fitting algorithm, the peak periods for Sector 22, 23, and 49 can be calculated as 0.54919346 d, 0.54945736 d, and 0.54942793 d respectively in Python. This alignment verifies the existence of the first exoplanet TOI-1807b in the TOI-1807 system as the peak period of $\sim$0.549 d is consistent with the published value of 0.549374 d \citet{2022arXiv220603496N}. To further detect a possible second transiting exoplanet in this system, we need to mask the transit signal of TOI-1807b and analyze the BLS spectra shown in Fig.~\ref{Fig4} after the masking process.

   Since the peak periods highlighted in thick blue lines in three periodograms in Fig.~\ref{Fig4} don’t agree with each other and these peak periods are not distinguished well from other fluctuations, hence there is no second transiting exoplanet with the periods from 0.5 d to 10.0 d in the TOI-1807 system. The maximum searching period in Fig.~\ref{Fig4} is set as 10 days because the choice of maximum period $P_{\mathrm{max}}$ for TESS is just 10 days by taking the cost of the TESS mission duration into account as well as the efficiency of transit detection \citep{2015JATIS...1a4003R}. Prior to the TESS mission initiation on 2018 April 18, \citet{2015ARA&A..53..409W} found and summarized that the occurrence rate of discovered exoplanets based on the Kepler mission was significantly decreased when the period is less than 10 days, which intrigues researchers to detect more exoplanets for $\emph{P\/} < 10$ days. Additionally, in both Fig.~\ref{Fig3} and Fig.~\ref{Fig4}, the BLS power fluctuates much more drastically in the periodograms of Sector 22 and Sector 23 compared with the fluctuation in the periodogram of Sector 49. This is mainly due to the artificial transits occurring when the camera didn’t record any stellar flux data, resulting in larger errors in the original light curves of Sector 22 and Sector 23. Although the drastic fluctuation may seriously affect the quality of the BLS spectra, it also reflects that the BLS spectrum of Sector 49 illustrates the most reliable result. Hence, we can confidently confirm that the second transiting exoplanet with periods varied from 0.5 d to 10.0 d doesn’t exist in the TOI-1807 system as the BLS power of Sector 49 fluctuates around a low value in Fig.~\ref{Fig4} with no obvious peak. Furthermore, we plotted the flattened, phase-folded light curves of Sector 22, 23, and 49 after masking the transit signal of TOI-1807b and we didn’t see any new transit signals by visual inspection, which is another strong evidence to rule out the existence of a second transiting exoplanet in the TOI-1807 system.

   There is indeed a more accurate method called the Transit Least Squares (TLS) analysis to better illustrate the peak periods in the periodograms. Instead of analyzing the BLS spectra, \citet{2022arXiv220603496N} plotted the TLS spectra of Sector 22 and Sector 23 in their paper to similarly detect a possible second transiting exoplanet with a wider range of periods in the TOI-1807 system after they masked the transit signal of TOI-1807b. It turns out that they obtained a weak peak period in the periodogram at $\sim$24.986 d, but they safely ruled out the possibility of another transit signal as they also couldn’t see any transit feature in their phase-folded light curves of Sector 22 and Sector 23 by visual inspection (which is compatible with our evidence aforementioned). If their findings are correct, we can further deduce that a second transiting exoplanet with a period less than 25 days is unlikely to exist in the TOI-1807 system. However, the existence of a second transiting exoplanet is still not clear when the period is larger than 25 days because inner USP planets are sometimes accompanied by their outer planets with longer periods, although the opportunity of detecting the transits of their outer planets is low (\citep{2018NewAR..83...37W, 2019MNRAS.488.3568P}). More specifically, \citet{2014ApJ...787...47S} found that the transit probability of detecting a USP planet is generally $\sim$7 times higher than the transit detection of its outer companions with periods ranging from 1 day to 50 days, and sometimes the outer planets are not even detectable through the transit method due to their larger orbital semi-major axis or the large mutual inclination between their misaligned non-coplanar orbital plane \citep{Adams_2017}, which increases the possibility of existing a second non-transiting exoplanet in the TOI-1807 system. This inconclusive question might be further resolved after the implementation of the James Webb Space Telescope (JWST) launched on 2021 December 25.

   \begin{figure*}[ht!]
   \centering
   \includegraphics[width=17.3cm]{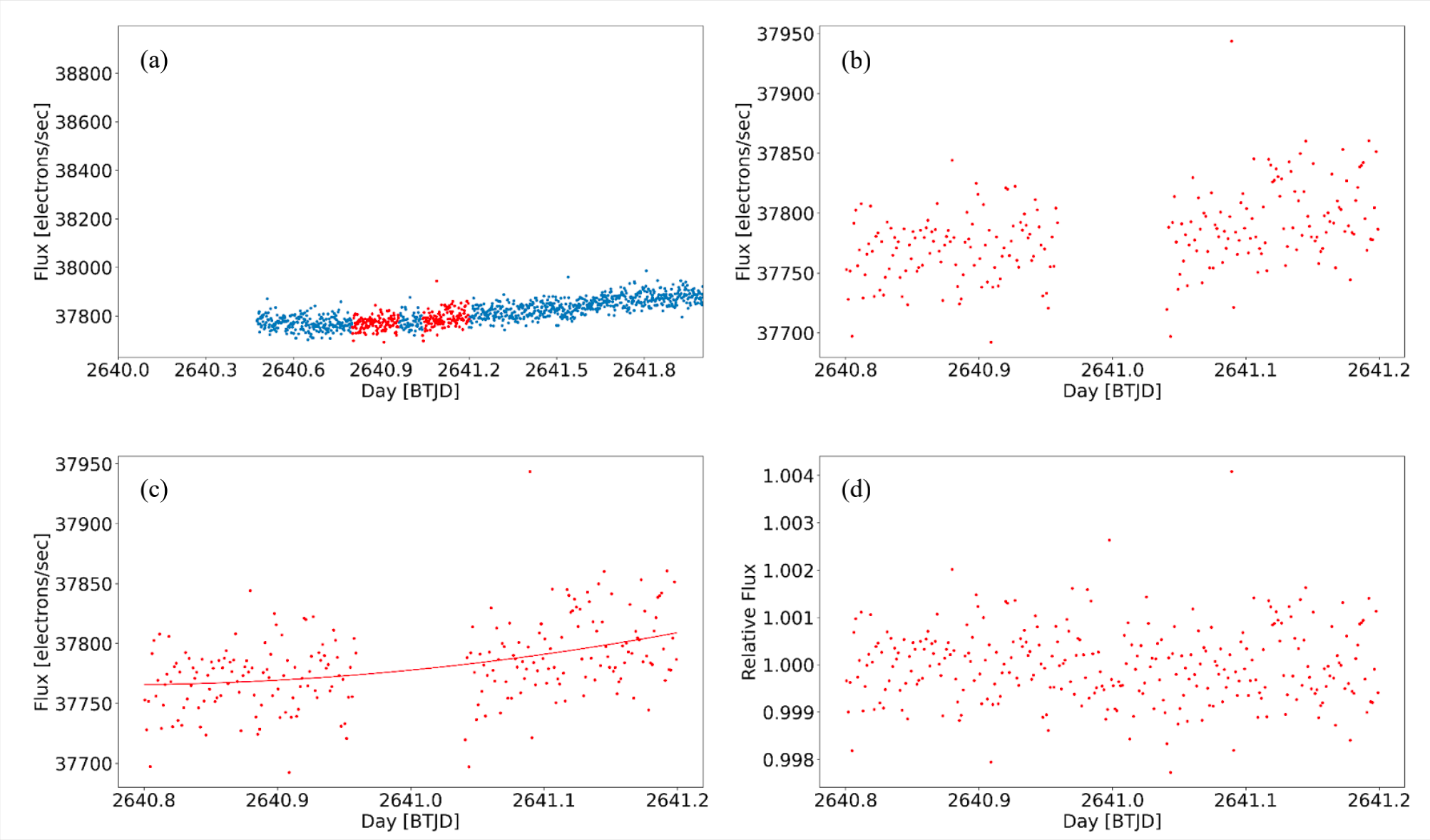}
   \caption{Fitting a selected range of Sector 49 data in each segment of the transit with polynomials.}
   \label{Fig5}
   \end{figure*}

   \begin{figure*}[ht!]
   \centering
   \includegraphics[width=17.3cm]{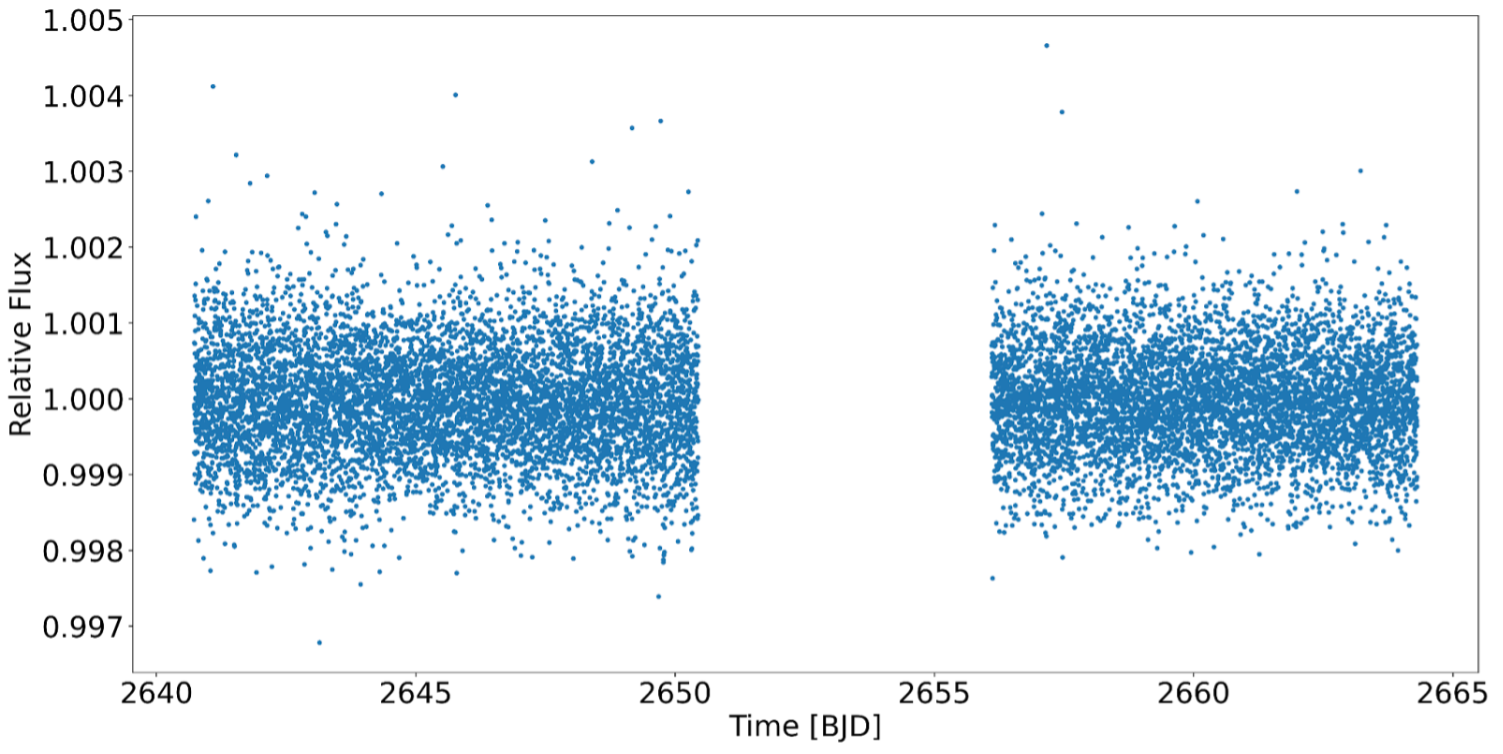}
   \caption{The local flattened light curve of TOI-1807 from TESS Sector 49 data.}
   \label{Fig6}
   \end{figure*}

\section{Methods}
\label{section3}

   The flattening code is a fundamental technique widely used in the analytical research of transits. By flattening the source code, we can remove long-term variations related to spacecraft motion and stellar fluctuations. In Section~\ref{section2}, the source code plot is presented in Fig.~\ref{Fig2}. The way that flattening code (or so-called Self Flat Fielding method, SFF) achieves its function of removing the spacecraft motion noise is by fitting the long-term trends with a low-order polynomial. There are several problems with this method: First, if we fit all the source data with polynomials, transit data points are included in the fit; Second, the overall variation in stellar noise may not be properly described by a single polynomial. Both problems will affect the accuracy of the flattened result. To fix these problems, we made our own SFF code called local flattening (see panel (a) of Fig.~\ref{Fig5}). We take the period estimation and fit each transit locally to remove local non-transit variations. The data outside of the transit region is fitted locally with a cubic polynomial, and then this process is repeated for all transits. The resulting Sector 49 light curve is in Fig.~\ref{Fig6}.

\begin{figure*}[ht!]
   \centering
   \includegraphics[width=17.3cm]{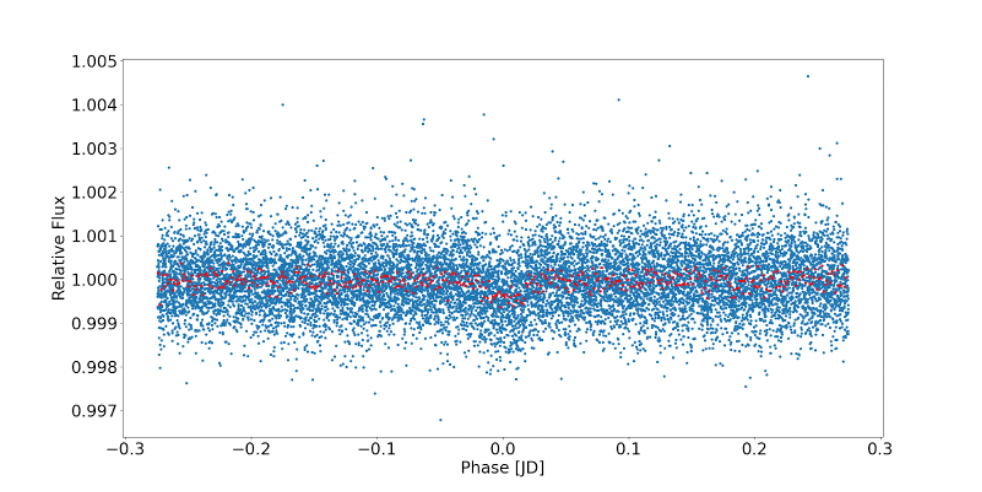}
   \caption{The flattened, phase-folded light curve of Sector 49 which is plotted for transit visualization. Each data point in red represents the averaged position of 30 data points in blue.}
   \label{Fig7}
   \end{figure*}

   After local flattening, we folded the transit data with a period equal to 0.549374 d to confirm the existence of transit. Comparing the phase-folded light curves of Sector 22, 23, and 49, a small transit dip is apparent around the central conjunction time, although the quality of the transit dip is seriously affected by other data points outside of the transit region with the values of relative flux distributed in a relatively wide range around 1. Hence, to keep illustrating the transit dip more distinctly, the averaging data method was used as a result. The aim of averaging data is to select the data each representing the mean position of a fixed number of other data points. These mean positions could better illustrate the shape of the light curve so that the transit dip becomes more apparent. The graph of the phase-folded light curve, as well as the averaged data points, are demonstrated in Fig.~\ref{Fig7}, where only the light curve of Sector 49 is included. The transit visualization in the light curves of Sector 22 and 23 are illustrated in Appendix~\ref{AppendixA}.
   
   To model the transit, we use the algorithm from \citet{2002ApJ...580L.171M}, {\fontfamily{pcr}\selectfont mandelagol}, to model the planetary transits. To simplify the motion of the orbiting planet during each transit, we first defined a phase of transit $\phi$ which varies linearly in time \emph{t\/}: 
   \begin{equation}
      \phi\left(t\right) = \frac{2\pi\left(t-t_{\mathrm{c}}\right)}{P}\,,\label{equation1}
   \end{equation}
   Where $t_{\mathrm{c}}$ is the conjunction time, and \emph{P\/} is the period of the planet. The phase of transit $\phi$ is periodic in 2$\pi$ because the planet completes a circle whenever \emph{t\/} – $t_{\mathrm{c}}$ = \emph{P\/}. Based on this definition, we could further construct a Cartesian coordinate system (\emph{X\/}, \emph{Y\/}, \emph{Z\/}) centered at the center of the star to better locate the relative position of the planet:
   \begin{eqnarray}
      X\left(t\right) = a\sin{\phi\left(t\right)}\,,\label{equation2}\\
      Y\left(t\right) = a\sin{i}\cos{\phi\left(t\right)}\,,\label{equation3}\\
      Z\left(t\right) = a\cos{i}\cos{\phi\left(t\right)}\,,\label{equation4}
   \end{eqnarray}
   Where \emph{i\/} is the inclination angle which is defined as the angle between the planetary orbit normal and the line of sight, \emph{a\/} = $a_{\mathrm{p}}$/$R_{\mathrm{\ast}}$ is the ratio of the orbital semi-major axis of the planet $a_{\mathrm{p}}$ to the stellar radius $R_{\mathrm{\ast}}$. If one knows the planetary radius $R_{\mathrm{p}}$, the size ratio \emph{k\/} = $R_{\mathrm{p}}$/$R_{\mathrm{\ast}}$ is also obtained. In the case of TOI-1807b, since the planetary semi-major axis ap is much larger than the stellar radius $R_{\mathrm{\ast}}$ and $\emph{k\/} \ll 0.1$, the planet’s motion during the transit can be assumed as a straight line \citep{2002ApJ...580L.171M}. Therefore, equation (\ref{equation3}) can be rewritten as the following expression:
   \begin{equation}
   Y\left(t\right) = b\cos{\phi\left(t\right)}\,,\label{equation5}
   \end{equation}
   Where $\emph{b\/} = \emph{a\/}\cos{i}$ represents the transit impact parameter. So far, we have defined the key equations in the {\fontfamily{pcr}\selectfont mandelagol} package from equation (\ref{equation1}) to (\ref{equation5}). The essential parameters consist of the conjunction time $t_{\mathrm{c}}$, the period of the planet \emph{P\/}, the size ratio \emph{k\/}, the ratio of the planetary semi-major axis to the stellar radius \emph{a\/}, the transit impact parameter \emph{b\/}, and the limb darkening coefficients $u_{\mathrm{1}}$, $u_{\mathrm{2}}$. After giving the initial guess of these essential parameters from the published value \citep{2022arXiv220603496N}, the flattened light curves of Sector 22, 23, and 49 can then be fitted. However, due to the lack of information, we decided not to fit two limb darkening coefficients but to quote the published value \citep{2022arXiv220603496N} to reduce the parameter uncertainties encountered in the later optimizing process. The transit impact parameter \emph{b\/} is strongly correlated with parameter \emph{a\/}, which means it is necessary to fit both two parameters to avoid underestimating the uncertainty of each parameter. Thus, there are five parameters analyzed in this paper.

   Since the initial guess is not accurate enough, we need to use the curve-fit code in Python to obtain the best-fit parameters as well as their uncertainties for all three sectors. One should notice that the fit of the folded flattened light curves is plotted only for transit visualization. While the parameters are optimized according to the {\fontfamily{pcr}\selectfont mandelagol} transit model as aforementioned, the calculation of their uncertainties is slightly more complicated, which requires estimating the uncertainty in the flux measurement of each data point. A quick estimation method is to first select and exclude the data points in the transit region as precisely as possible. For the rest of the data points outside of the transit region, the standard deviation is calculated and assumed as the constant uncertainty for all the data points. The uncertainties in the best-fit parameters can therefore be gauged after running through the curve-fit code. Finally, a hypothesis test including both the reduced $\chi^2$ and the P-value is performed to evaluate whether the transit model with its best-fit parameters is reasonable enough to fit the flattened light curves of Sector 22, 23, and 49.

   However, the uncertainties in best-fit parameters are still not reliable because of the rough method to calculate the uncertainty in each data point. Hence, a computational algorithm called the affined invariant Markov Chain Monte Carlo (MCMC) with the sampler {\fontfamily{pcr}\selectfont emcee} is implemented to compute a more robust uncertainty. Thanks to the help of MCMC, we can randomly draw samples from the posterior probability distribution of parameters where each step in the Markov Chain only depends on the previous step \citep{Ravenzwaaij_2018, 2019Ap&SS.364...33H}. In addition, MCMC can also help us remove the nuisance parameters, which is known as the marginalization process \citep{Foreman_Mackey_2013}. After running through MCMC with 200,000 iterations, the eventual model describes the highest likelihood of the data, meaning that it fits the data reasonably well. The curve-fit parameters as the initial guess of MCMC are further optimized as a result. In order to obtain this highest likelihood model, it is necessary to define the likelihood function and the prior function to constrain the parameter advances simulated by the MCMC algorithm. The likelihood function is determined through a $\chi^2$ test where only the parameters passing this test can be kept by the MCMC algorithm to adjust the parameter sets:
   \begin{equation}
   P\left(Y|\theta\right) = \frac{1}{2}\sum\left(\frac{Y_{\mathrm{data}}-Y_{\mathrm{model}}}{Y_{\mathrm{error}}}\right)^2\,,\label{equation6}
   \end{equation}
   Where \emph{Y\/} is the dependent variable in a model function, $\theta$ is a parameter vector used to generate the highest likelihood model. In this paper, it is expressed in a 5-tuple:
   \begin{equation}
   \theta = \left( \begin{array}{c} t_{\mathrm{c}} \\ \emph{P\/} \\ \emph{k\/} \\ \emph{a\/} \\ \emph{b\/} \end{array}\right)\,,\label{equation7}
   \end{equation}
   The prior function is then applied to set the initial conditions for the parameter vector $\theta$:
   \begin{equation}
   P\left(\theta\right) = \left\{ \begin{array}{l}
   1000 < t_{\mathrm{c}} < 4000\,, \\ 0.5 < \emph{P\/} < 0.6\,, \\
   0.015 < \emph{k\/} < 0.020\,, \\ 1 < \emph{a\/} < 10\,, \\ 0 < \emph{b\/} < 1\,,
   \end{array}\right.\label{equation8}
   \end{equation}
   
   The range of each parameter is adopted to be relatively wider so that any potential errors from the curve-fit parameters will be reviewed by the MCMC algorithm to obtain a more precise result. Based on the definition of the likelihood function $P\left(Y|\theta\right)$ and the prior function $P\left(\theta\right)$, we can derive the posterior probability distribution of parameters which MCMC draws samples from according to Bayes’ Theorem \citep{Foreman_Mackey_2013}:
   \begin{equation}
   P\left(\theta|Y\right) \sim P\left(Y|\theta\right)P\left(\theta\right)\,,\label{equation9}
   \end{equation}
   Similarly, the reduced $\chi^2$ and P-value hypothesis tests are performed again to analyze the reasonability of the highest likelihood model obtained by MCMC. A comparison is made between the curve-fit parameters and the MCMC parameters for Sector 22, 23, and 49 to prove that MCMC is indeed a more robust method with more reliable uncertainties. The light curves of three sectors fitted by the {\fontfamily{pcr}\selectfont mandelagol} transit model as well as their normalized residuals are plotted in Python, and artificial transits mentioned in Section~\ref{section2} are clipped away to reduce the error.

\section{Results}
\label{section4}

\begin{figure*}[ht!]
   \centering
   \includegraphics[width=17.3cm]{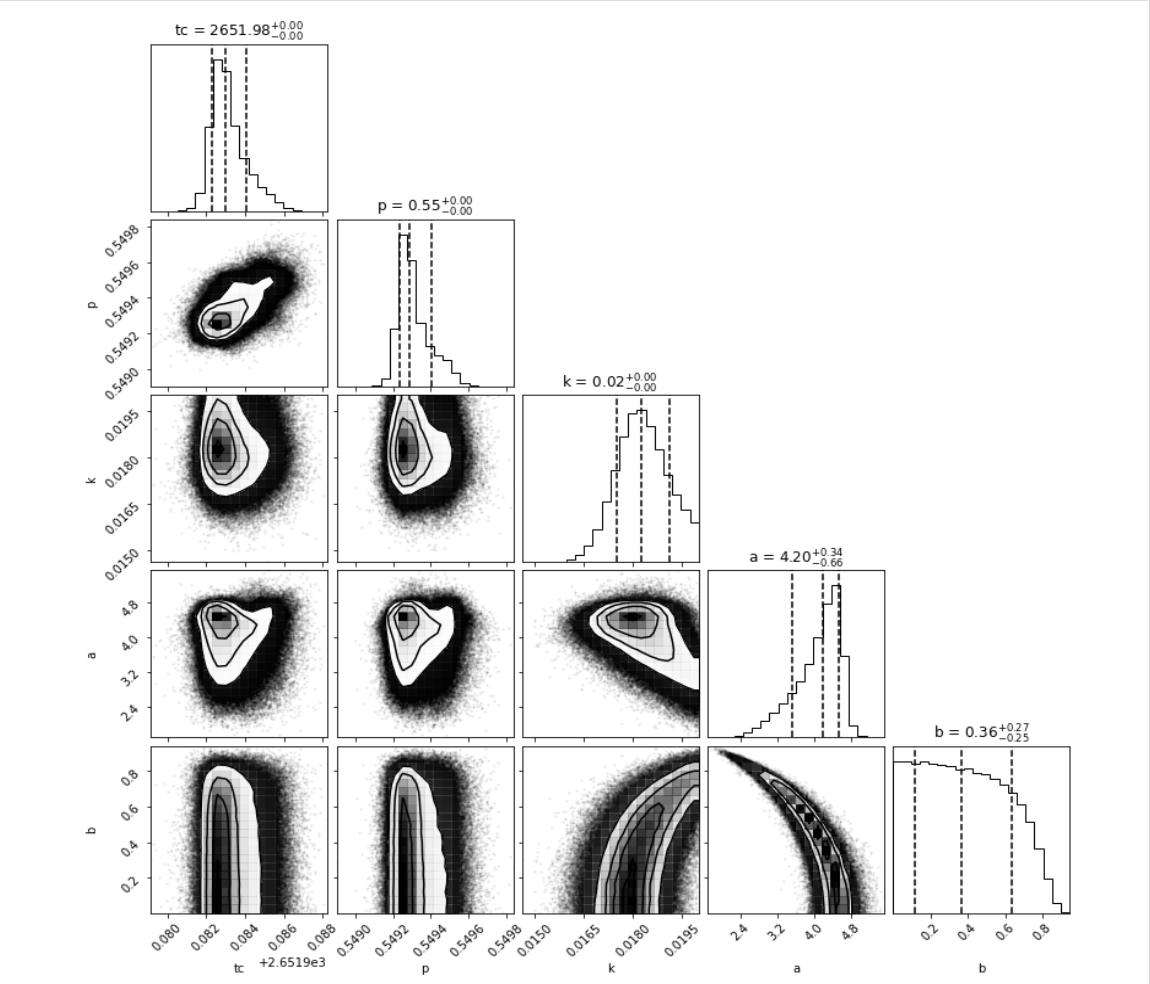}
   \caption{The corner plot of Sector 49 parameters obtained from the MCMC algorithm.}
   \label{Fig8}
   \end{figure*}

   The results of parameter fitting via the MCMC algorithm are demonstrated in Table~\ref{Table1}, Table~\ref{TableC1}, and Table~\ref{TableC2}. Using the methods described in Section~\ref{section3}, preliminary curve-fit optimization is done with Python’s {\fontfamily{pcr}\selectfont SciPy} module, giving us the initial values for the MCMC algorithm, in which an uninformative prior is used. The MCMC results include parameters of maximum likelihood as well as the $\pm\sigma$ errors. The stellar limb darkening coefficient was taken to be quadratic with $u_{\mathrm{1}}$ = 0.46, $u_{\mathrm{2}}$ = 0.17 determined by \citet{2022arXiv220603496N}, and hence not considered in the MCMC simulation. The limb darkening coefficients are usually determined judging by the shape of the “transition” between ingress/egress and full transit. The short duration of transits of USP planets would also make it difficult to constrain the limb darkening coefficients. These may be better determined with reference to specific stellar structures. The impact parameter is further correlated to the orbital semi-major axis of the USP, as shown in Fig.~\ref{Fig8}. The corner plot of MCMC fitting to Sector 49 with 5 parameters shows that parameter \emph{a\/} is strongly affected by the changes in the impact parameter \emph{b\/} which is only slightly excluded at $\sim$1, but we include \emph{b\/} for fitting to accurately gauge the errors on the orbital semi-major axis. To check the quality of the parameters, a phase folded light curve is produced with the period and conjunction times given in Table~\ref{Table1}, using a $\chi^2$ test. The parameter results agree reasonably well across all three sectors. Note that the value for $t_{\mathrm{c}}$ is chosen differently for Sector 49 than Sector 22 and 23. Sector 22 and 23 cover data from BTJD 1900 to 1955, while Sector 49 covers more recent data from BTJD 2640 to 2665. It was discovered that our initial value for $t_{\mathrm{c}}$ = 1899.3449 BTJD resulted in the Monte Carlo simulation failing to converge. A more reasonable estimate was chosen for Sector 49 at 2651.987 BTJD (as opposed to 1899.3449 BTJD used in previous simulations) in the middle of the sector to reduce any effect our initial guess may have had on the parameter optimization.

   The MCMC simulation was completed once we judged the contours to be smooth and converged. The model best fit is plotted for the phase-folded light curve to judge the quality of the fit. The fits and corner plots for Sector 49 are only shown in Fig.~\ref{Fig8} and Fig.~\ref{Fig9}, while the analysis of Sector 22 and 23 are included in Appendix~\ref{AppendixB} and~\ref{AppendixC}. The plots demonstrate that the transit shape and timing are very well described by the parameters fitted. All the transit light curves are centered at zero, which suggests that the conjunction time is computed correctly. In addition, all the normalized residual plots show that errors are evenly distributed about the central line (have a mean of zero), and 96\% are contained within $\pm2$, which indicates a good fit \citep{Hughes_2010}. Since no obvious trends (i.e., a “bump” reveals an unexplained transit) exist in the normalized residual plot, it also suggests that the data are consistent with the transit fitting model \citep{Hughes_2010}.

   Furthermore, we performed a hypothesis test by computing the reduced $\chi^2$ value and its corresponding P-value to evaluate the statistical power of fits. Sector 22 has reduced $\chi^2$ 1.03 with a P-value of 0.002 owing to the large systematic error in the data, Sector 23 has reduced $\chi^2$ 1.022 and a P-value of 0.05, which we deem to be acceptable given the noise present in the data. Sector 49 yields the best statistics with reduced $\chi^2$ 1.002 and a P-value of 0.43. The hypothesis test results of the three sectors all indicate that the transit model is reasonable enough to fit the data, while in contrast, the fit to the data of Sector 49 is the most reasonable fit among three sectors \citep{Hughes_2010}.

   \begin{figure*}[ht!]
   \centering
   \includegraphics[width=17.3cm]{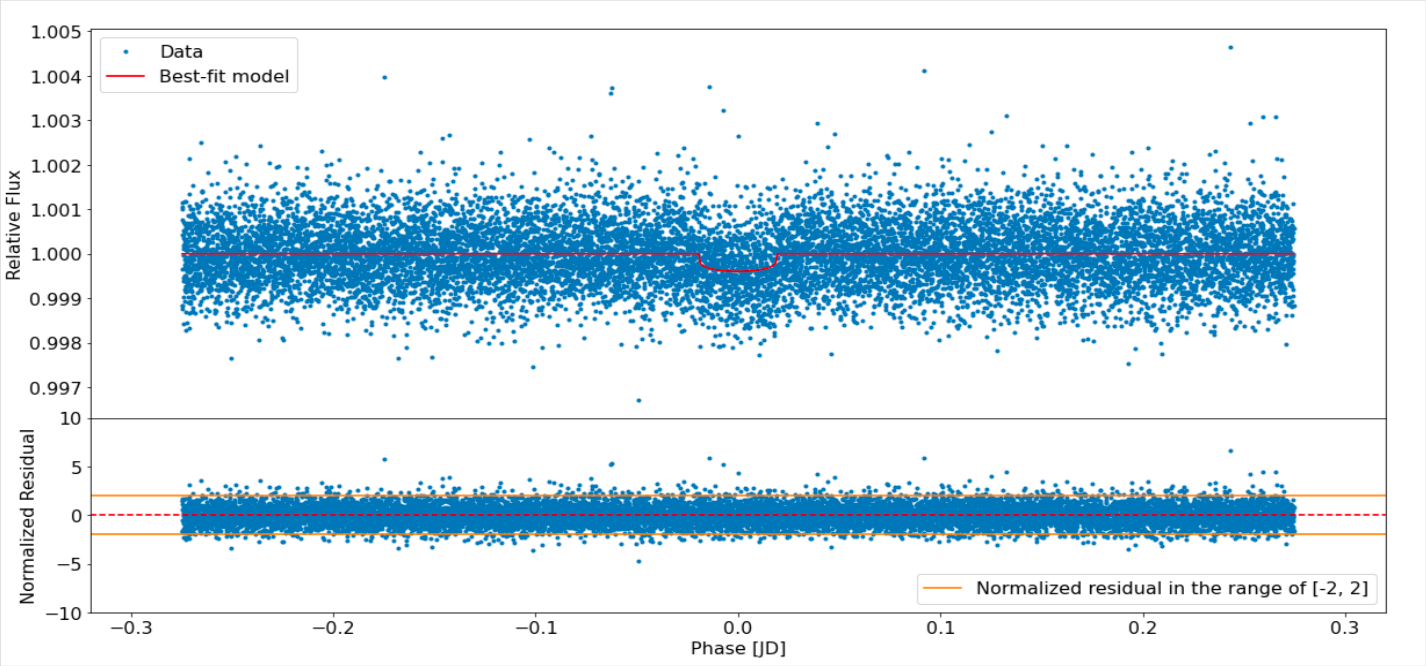}
   \caption{The light curve fitted by the transit model as well as the normalized residual plot for Sector 49.}
   \label{Fig9}
   \end{figure*}

\section{Discussion}
\label{section5}
\subsection{The surface temperature, mass, radius, and density of TOI-1807b}
\label{section5.1}
   
   Due to the reason that the data from Sector 49 have never been analyzed in any published papers before, we chose the parameter fitting result from Sector 49 to theoretically calculate the mass, radius, and density of TOI-1807b and compare our results to the published value \citep{2022arXiv220603496N}. The summarized results are organized in Table~\ref{Table2}. To begin with, we introduce Newton-Kepler’s formula:
   \begin{equation}
   \frac{M_{\mathrm{p}}}{M_{\mathrm{\ast}}} = \frac{a_{\mathrm{\ast}}}{a_{\mathrm{p}}}\,,\label{equation10}
   \end{equation}
   Where $M_{\mathrm{p}}$ is the mass of the planet, $M_{\mathrm{\ast}}$ is the mass of the star, and $a_{\mathrm{\ast}}$ is the orbital semi-major axis of the star. By using Newton-Kepler’s formula, we assumed that there is only one planet in this planetary system (which so far is true). Thus, we need to know the remaining three parameters: $M_{\mathrm{\ast}}$, $a_{\mathrm{\ast}}$, $a_{\mathrm{p}}$. The mass of TOI-1807 is $M_{\mathrm{\ast}} = 0.76 \pm 0.03\, M_\odot$ given by \citet{2022arXiv220603496N}. Meanwhile, we take the value of the MCMC parameter \emph{a\/} = $a_{\mathrm{p}}$/$R_{\mathrm{\ast}} = 4.20^{+0.33}_{-0.66}$ shown in Table~\ref{Table1}, and we can obtain $a_{\mathrm{p}} = 2.90^{+0.28}_{-0.48}\, R_\odot$ by assuming that $R_{\mathrm{\ast}} = 0.690 \pm 0.036\, R_\odot$ \citep{2022arXiv220603496N}. Then, we must introduce another formula to proceed with the calculation of the mass of the planet $M_{\mathrm{p}}$:
   \begin{equation}
   v = \frac{2\pi a_{\mathrm{\ast}}}{P}\,,\label{equation11}
   \end{equation}
   Where $v = 2.39^{+0.45}_{-0.46}\, \mathrm{m\, s^{-1}}$ is the relative radius speed of the star \citep{2022arXiv220603496N} and \emph{P\/} is the period (we take the value of $P = 0.54929^{+0.00012}_{-0.00005}$ days from Table~\ref{Table1}). From this formula, $a_{\mathrm{\ast}} = 18.1^{+3.4}_{-3.5}$ km is can be computed. Eventually, we are able to use $M_{\mathrm{\ast}}$, $a_{\mathrm{\ast}}$, and $a_{\mathrm{p}}$ to calculate the mass of TOI-1807b as $M_{\mathrm{p}} = 2.27^{+0.49}_{-0.58}\, M_\oplus$, which gives a percentage uncertainty of 25.6\%. In addition, we also take the value of the MCMC parameter \emph{k\/} = $R_{\mathrm{p}}/R_{\mathrm{\ast}} = 0.01825^{+0.00086}_{-0.00076}$ to calculate the radius of TOI-1807b as $R_{\mathrm{p}} = 1.37^{+0.10}_{-0.09}\, R_\oplus$, giving a percentage uncertainty of 7.03\%. The density of TOI-1807b is therefore $\rho_{\mathrm{p}} = 0.875^{+0.264}_{-0.285}\, \rho_\oplus$ with a percentage uncertainty of 32.5\%. All these measurements agree within their $\pm1\sigma$ errors. This new value of Earth-like density infers that the current composition of TOI-1807b might be consistent with pure silicate rock, or the composition with a less massive iron core surrounded by a more massive silicate rock shell and other less dense volatiles such as water \citep{2018NewAR..83...37W} and other volatile elements such as H, C, N, S, and Cl \citep{2009ApJ...703L.113S}, although in the future TOI-1807b might experience an iron enhancement process due to sublimated silicates evaporation driven by a hydrodynamic wind \citep{2013MNRAS.433.2294P, 2020ApJ...894....8P}, which makes the composition of TOI-1807b gradually transfer from Earth-like composition (70\% rock and 30\% iron) to Mercury-like composition (30\% rock and 70\% iron).

   To calculate the surface temperature of TOI-1807b, we begin to set up the scheme by taking the star as an ideal blackbody radiation source so that we can use the Stefan-Boltzmann law to calculate the stellar luminosity $L_{\mathrm{\ast}}$:
   \begin{equation}
   L_{\mathrm{\ast}} = 4\pi R_{\mathrm{\ast}}^2\sigma T_{\mathrm{\ast}}^4\,,\label{equation12}
   \end{equation}
   Where $\sigma$ is the Stefan-Boltzmann constant, and $T_{\mathrm{\ast}}$ is the surface temperature of the star. While the planet, as a power receiver, can only absorb the power irradiating its surface:
   \begin{equation}
   P = L_{\mathrm{\ast}}\frac{\pi R_{\mathrm{p}}^2}{4\pi a_{\mathrm{p}}^2}\,,\label{equation13}
   \end{equation}
   In this formula, the albedo of the planet is taken to be 0, which means all the starlight irradiating the planet will be absorbed. However, most planets have an albedo other than 0, hence we assume the value of albedo as the bond albedo of Earth $A_{\mathrm{B}}$ = 0.306 since all confirmed USP planets from NASA Exoplanet Archive had taken this value as their albedo \citep{Hedges_2021}. When the power the planet absorbed is equal to the power the planet radiated, thermal equilibrium is attained ($P_{\mathrm{in}}$ = $P_{\mathrm{out}}$) and we can obtain the following equation if we suppose that the temperature is constantly distributed over the surface of the planet:
   \begin{eqnarray}
      P_{\mathrm{in}} = \left(1-A_{\mathrm{B}}\right)L_{\mathrm{\ast}}\frac{\pi R_{\mathrm{p}}^2}{4\pi a_{\mathrm{p}}^2}\,,\label{equation14}\\
      P_{\mathrm{out}} = 4\pi R_{\mathrm{p}}^2\sigma T_{\mathrm{p}}^4\,,\label{equation15}
   \end{eqnarray}
   The expression for $T_{\mathrm{p}}$ is then derived as:
   \begin{equation}
   T_{\mathrm{p}} = T_{\mathrm{\ast}}\left(\frac{R_{\mathrm{\ast}}}{2a_{\mathrm{p}}}\right)^\frac{1}{2}\left(1-A_{\mathrm{B}}\right)^\frac{1}{4}\,,\label{equation16}
   \end{equation}
   Where $T_{\mathrm{\ast}} = 4757^{+51}_{-50}$ K \citep{Hedges_2021}, $R_{\mathrm{\ast}} = 0.690 \pm 0.036\, R_\odot$, and $a_{\mathrm{p}} = 2.90^{+0.28}_{-0.48}\, R_\odot$. Substituting these values and their uncertainties into equation (\ref{equation16}), we can obtain a much lower planet’s equilibrium temperature of $T_{\mathrm{p}} = 1499^{+82}_{-129}$ K different from $T_{\mathrm{p}} = 2100^{+39}_{-40}$ K published by \citet{Hedges_2021}. This is attributed to the relatively larger value of the orbital semi-major axis we obtained in this paper compared with the previous value \citep{Hedges_2021}, where ${a_{\mathrm{p}}}^{-1/2}$ is proportional to $T_{\mathrm{p}}$. Even though this correction computes a more reliable result, the new equilibrium temperature of TOI-1807b might still be higher than the actual situation because the temperature can’t be constant everywhere over the surface of the planet. The surface of the planet is always divided into dayside surface and nightside surface, where the nightside temperature is significantly cooler than the dayside temperature.

   Besides, one may also be curious about the possible atmosphere existing over the surface of TOI-1807b. For most USP planets, their atmospheres are very thin or have already evaporated off into space due to the long-term effects of high-energy irradiation (such as ultraviolet and X-rays radiation) emitted from the extremely active host stars in their early phase \citep{2018NewAR..83...37W}. The USP planets are orbiting so close to their host stars, which means their outer atmospheres are heated so intensively that the molecules of atmospheres can easily experience the thermal oscillation and be accelerated to exceed their escape velocity before they are no longer pulled by the planet’s gravitational field, even though most USP planets have the protection of their magnetic field to prevent the molecules from escaping. Meanwhile, a Parker wind generated by both the stellar continuum radiation and the planet’s internal heat will intensify the mass loss of the hydrogen-helium envelope atmosphere \citep{1958ApJ...128..664P, 2016ApJ...817..107O}, while other cases, including another wind generated by the pressure gradient instead in an extremely heating atmospheric environment of early USP planets and the photoevaporation process, can even cause the complete mass loss of any hydrogen-helium envelope atmosphere when the host stars are still active in their youth \citep{2013ApJ...776....2L, 2014ApJ...787...47S, 2018NewAR..83...37W}. Although the new equilibrium temperature of TOI-1807b has not reached the melting point of iron, it is higher enough to melt or even sublime most common metals, nonmetals, and solid compounds (O, $\ce{O2}$, Mg, SiO, $\ce{SiO2}$, etc.) in the crust and mantle of TOI-1807b. Therefore, the surface of TOI-1807b is not only a “lava world”, but it is also a world with a possible thin layer of atmosphere mainly composed of $\ce{N2}$, $\ce{O2}$, O, and SiO, where the surface temperature is in the range of 1500--3000 K \citep{2009ApJ...703L.113S, 2018NewAR..83...37W}. It is unlikely that the atmosphere of TOI-1807b is completely lost because the sublimated silicates evaporation mentioned above is highly possible to be ongoing, which means TOI-1807b might be currently experiencing its iron enhancement process \citep{2020ApJ...894....8P}.

\subsection{Orbital decay}
\label{section5.2}

   From the parameters obtained in the results section, we can further explore various properties of our target exoplanet TOI-1807b. Firstly we investigate the possible period shrinkage that might occur in the orbit of TOI-1807b. \citet{2017AJ....154....4P} stated that both tidal orbital decay and apsidal precession could affect the period measurement because sometimes apsidal precession is falsely regarded as orbital decay. The orbits of hot Jupiters have long been predicted to shrink due to tidal orbital decay \citep{1996ApJ...470.1187R, 2009ApJ...692L...9L, 2017AJ....154....4P}. Although TOI-1807b is an Earth-like rocky planet rather than a hot Jupiter, \citet{2018NewAR..83...37W} also predicted that the USP planets are very likely to experience an ongoing orbital decay. In a planetary system consisting of only one planet whose orbit is circular (\emph{e\/} = 0), the tidal evolution is not stable when the orbital angular momentum of the planet is more than three times the stellar spin angular momentum of the star \citep{1980A&A....92..167H}, which results in the orbital shrinkage of the planet. In addition, the orbital decay of the USP planets can also be explained by the tidal bulge theory. Due to the gravitational force between the star and the planet, the shape of the planet is deformed along the axis toward the star, and the star forms a tidal bulge along the axis toward the planet as well. Since the orbital period of USP planets ($\emph{P\/} < 1$ day) is much smaller than the rotational period of early-time Sun-like stars ($\sim$10 days), the lagged tidal bulge of the star will sap the orbital angular momentum of the USP planets, leading to the consequence of orbital decay. For our TOI-1807 system, the ratio of the stellar spin angular momentum $L_{\mathrm{spin}}$ to the orbital angular momentum $L_{\mathrm{orbit}}$ is determined by \citet{2019MNRAS.488.3568P}:
   \begin{equation}
   \begin{aligned}
   \frac{L_{\mathrm{spin}}}{L_{\mathrm{orbit}}} = 35\left(\frac{k_{\mathrm{\ast}}}{0.06}\right)\left(\frac{M_{\mathrm{p}}}{M_{\mathrm{\oplus}}}\right)^{-1}\left(\frac{a_{\mathrm{p}}}{0.02\, AU}\right)^{-\frac{1}{2}}\\
   \times\left(\frac{M_{\mathrm{\ast}}}{M_{\mathrm{\odot}}}\right)^{\frac{1}{2}}\left(\frac{R_{\mathrm{\ast}}}{R_{\mathrm{\odot}}}\right)^{2}\left(\frac{P_{\mathrm{\ast}}}{30\, d}\right)^{-1}\,,\label{equation17}
   \end{aligned}
   \end{equation}
   Where $k_{\mathrm{\ast}}$ is a constant defined by the moment of inertia of TOI-1807 and is approximately equal to 0.06 for a Sun-like star \citep{2019MNRAS.488.3568P}, $P_{\mathrm{\ast}} = 8.83 \pm 0.08$ days is the stellar rotational period of TOI-1807 \citep{2022arXiv220603496N}. The ratio yields a value of $26.5^{+6.5}_{-7.7}$ where we take $M_{\mathrm{p}} = 2.27^{+0.49}_{-0.58}\, M_\oplus$, $a_{\mathrm{p}} = 0.0135^{+0.0013}_{-0.0022}$ AU, $M_{\mathrm{\ast}} = 0.76 \pm 0.03\, M_\odot$, and $R_{\mathrm{\ast}} = 0.690 \pm 0.036\, R_\odot$ from the planetary parameters of Sector 49. This result doesn’t meet the mentioned stability criteria \citep{1980A&A....92..167H}, indicating that the orbital decay process of TOI-1807b might be very slow since the stellar spin angular momentum is dominated in the TOI-1807 system.

   \begin{figure}[ht!]
   \centering
   \includegraphics{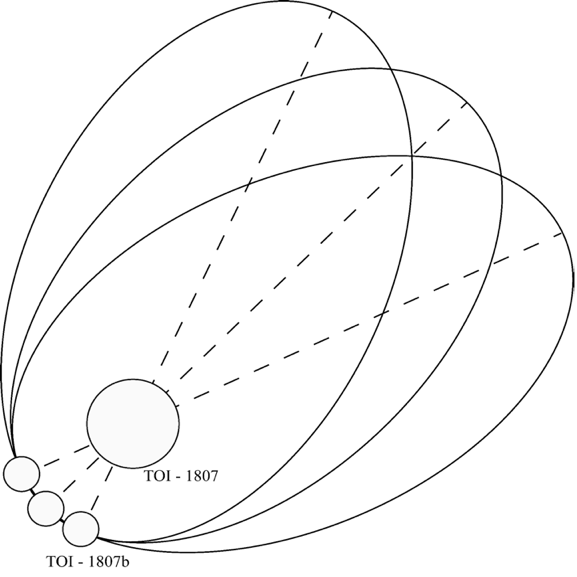}
   \caption{The exaggerated apsidal precession diagram of TOI-1807b when it is orbiting around its host star TOI-1807.}
   \label{Fig10}
   \end{figure}

   When it comes to apsidal precession, it is closely related to the planet’s density distribution \citep{2017AJ....154....4P}. Considering different terms contributing to the planet’s apsidal precession, it is generally dominated by the planetary tidal bulge \citep{2019AJ....157..180P}, thus the expression to calculate the apsidal precession rate of the USP planets only includes the term due to the planet’s tidal deformability. For our case of TOI-1807b, the expression can be derived and rewritten as \citep{1939MNRAS..99..451S, 2009ApJ...698.1778R, 2019MNRAS.488.3568P}:
   \begin{equation}
   \begin{aligned}
   \dot{\omega} = \left(2.32\times10^{-12}\, rad/s\right)\left(\frac{k_{\mathrm{p}}}{0.3}\right)\left(\frac{M_{\mathrm{\ast}}}{M_{\mathrm{\odot}}}\right)^{\frac{3}{2}}\\
   \times\left(\frac{M_{\mathrm{p}}}{M_{\mathrm{\oplus}}}\right)^{-1}\left(\frac{R_{\mathrm{p}}}{R_{\mathrm{\oplus}}}\right)^{5}\left(\frac{a_{\mathrm{p}}}{0.02\, AU}\right)^{-\frac{13}{2}}\,,\label{equation18}
   \end{aligned}
   \end{equation}
   Where $k_{\mathrm{p}}$ is the planet’s Love number. In this case, we assume $k_{\mathrm{\ast}} \approx 2$ for an Earth-like rocky planet \citep{Yoder95}. After substituting the published value \citep{2022arXiv220603496N} $M_{\mathrm{\ast}} = 0.76 \pm 0.03\, M_\odot$ and the values we obtained in this paper including $M_{\mathrm{p}} = 2.27^{+0.49}_{-0.58}\, M_\oplus$, $R_{\mathrm{p}} = 1.37^{+0.10}_{-0.09}\, R_\oplus$, and $a_{\mathrm{p}} = 0.0135^{+0.0013}_{-0.0022}$ AU into equation (\ref{equation18}), we can calculate the apsidal precession rate of $d\omega_{\mathrm{p}}/dt = 0.522^{+0.233}_{-0.238}\, \mathrm{deg\, yr^{-1}}$, which could hardly affect the orbital period of TOI-1807b unless the observation time is spanned for several decades (the current observation span in this paper is only 2 years). The apsidal precession diagram demonstrated in Fig.~\ref{Fig10} is exaggerated as a result. Hence, neither tidal orbital decay nor apsidal precession can trigger period shrinkage for TOI-1087b. Notwithstanding, it is still possible to calculate the timescale for the planet’s decay rate given by the following equation \citep{1966Icar....5..375G, Yee_2019, 2020AJ....159..150P}:
   \begin{equation}
       \dot{P} = -\frac{27\pi}{2Q_{\mathrm{\ast}}'}\left(\frac{M_{\mathrm{p}}}{M_{\mathrm{\odot}}}\right)\left(\frac{R_{\mathrm{\ast}}}{a_{\mathrm{p}}}\right)^5\,,\label{equation19}
   \end{equation}
   Where $Q_{\mathrm{\ast}}'$ is the star’s modified tidal quality factor, which was determined to be $10^{5.5}\sim10^{6.5}$ according to the research results found by \citet{Jackson_2008}, while \citet{2010ApJ...724L..53S} showed that their observed results better matched a larger range of value $10^{6}\sim10^{7}$ favored by both close-in giants and very hot Super-Earths. We presume a nominal value of $Q_{\mathrm{\ast}}' = 10^{6}$ for a moderate tidal evolution in this paper \citep{2010ApJ...724L..53S, 2020AJ....159..150P}, then we can calculate the decay rate of $dP/dt = - \left(9.20^{+5.41}_{-8.30}\right) \times 10^{-6}\, \mathrm{s\, yr^{-1}}$, revealing that it is nearly impossible to detect this change with current equipment precision. Furthermore, we can derive another timescale for the planet’s orbit shrinking to zero after knowing the current decay rate:

    \renewcommand{\arraystretch}{1.2}
   \begin{table*}[ht!]
        \begin{tabular}{p{7cm} p{1cm} p{3cm} p{3cm}}
        \hline\hline
        Parameter & Unit & MCMC result & Optimal value \\
        \hline
            Conjunction time ($t_{\mathrm{c}}$) & BTJD & $2651.98224^{+0.00112}_{-0.00064}$ & 2651.98224  \\
            Period (\emph{P\/}) & days & $0.54929^{+0.00012}_{-0.00005}$ & 0.54923 \\
            Planetary radius to stellar radius ratio (\emph{k\/}) & / & $0.01825^{+0.00086}_{-0.00076}$ & 0.0180 \\
            Orbital semi-major axis to stellar radius ratio (\emph{a\/}) & / & $4.20^{+0.33}_{-0.66}$ & 4.60  \\
            Transit impact parameter (\emph{b\/}) & / & $0.36^{+0.27}_{-0.25}$ & / \\
        \hline
        \end{tabular}
        \caption{The parameters of Sector 49 obtained from both the curve fit method and the MCMC algorithm.}
        \label{Table1}
   \end{table*}

   \begin{table*}[ht!]
        \begin{tabular}{p{7cm} p{1cm} p{3cm} p{3.2cm}}
        \hline\hline
        Parameter & Unit & Calculated result & Percentage uncertainty \\
        \hline
            Planetary mass ($M_{\mathrm{p}}$) & $M_{\mathrm{\oplus}}$ & $2.27^{+0.49}_{-0.58}$ & 25.6\%   \\
            Planetary radius ($R_{\mathrm{p}}$) & $R_{\mathrm{\oplus}}$ & $1.37^{+0.10}_{-0.09}$ & 7.03\% \\
            Planetary density ($\rho_{\mathrm{p}}$) & $\rho_{\mathrm{\oplus}}$ & $0.875^{+0.264}_{-0.285}$ & 32.5\% \\
            Surface temperature ($T_{\mathrm{p}}$) & K & $1499^{+82}_{-129}$ & 8.64\%  \\
            Orbital semi-major axis ($a_{\mathrm{p}}$) & AU & $0.0135^{+0.0013}_{-0.0022}$ & 16.5\% \\
            Inclination angle (\emph{i\/}) & deg & $85.0^{+3.8}_{-3.5}$ & 4.47\% \\
        \hline
        \end{tabular}
        \caption{The planetary parameters of TOI-1807b computed from Newton-Kepler’s formula as well as the TESS Sector 49 data.}
        \label{Table2}
   \end{table*}
   
   \begin{equation}
       \frac{P}{\dot{P}} = 5.16^{+3.03}_{-4.65}\, Gyr\,,\label{equation20}
   \end{equation}
   Given that the host star TOI-1807 is a younger star with an age of only $300 \pm 80$ Myr \citep{2022arXiv220603496N}, it is obvious that the planet still needs to experience a long time before its destruction, which also confirms that the orbital decay of TOI-1807b is currently not observational.

\subsection{Circular orbit}
\label{section5.3}

   So far, there is still a heated debate about how the USP planets get so close to their host stars, and it is unlikely that the USP planets initially stabilized their current orbits when they first formed. The most widely accepted explanation for the origins of the USP planets is that they experienced an inward migration due to tidal dissipation \citep{2020ApJ...905...71M} generated from either stellar tides or planetary tides. While \citet{2017ApJ...842...40L} agreed more about stellar tidal dissipation which caused the USP planets to migrate inwards from the position near the innermost edge of the protoplanetary disk where they first formed, \citet{2010ApJ...724L..53S} proposed that planetary tidal dissipation is the main source and the USP planets were undergoing migrations by dynamical interactions in the multi-planet systems. Meanwhile, the initial conditions of planet-driven tidal dissipation are different. \citet{2019AJ....157..180P} postulated a high-eccentricity migration scenario where the USP planets formed their high-eccentricity orbit by chaotic secular interactions before migrating inwards, although \citet{2019MNRAS.488.3568P} investigated the scenario of low-eccentricity migration which is also feasible for the USP planets to evolve. Despite so many possible theories to explain the inward migration of the USP planets due to tidal dissipation, they all reach a consensus that tidal dissipation could lead to further circularization of the orbits of the USP planets. Based on equation (25) of \citet{1966Icar....5..375G}, \citet{2017AJ....154....4P} derived an equation to estimate the timescale for tidal orbital circularization:
   \begin{equation}
       \tau = \frac{e}{\dot{e}} = \frac{2Q_{\mathrm{p}}}{63\pi}\left(\frac{M_{\mathrm{p}}}{M_{\mathrm{\ast}}}\right)\left(\frac{a_{\mathrm{p}}}{R_{\mathrm{p}}}\right)^{5}P\,,\label{equation21}
   \end{equation}
   Where $Q_{\mathrm{p}}$ is the planet’s tidal quality factor. Since our target exoplanet TOI-1807b is a terrestrial planet and all the terrestrial planets in the solar system have values of $Q_{\mathrm{p}}$ ranging from 10 to 190 \citep{1966Icar....5..375G}, we take an estimate of $Q_{\mathrm{p}}$ as 100. By substituting the published value \citep{2022arXiv220603496N} $M_{\mathrm{\ast}} = 0.76 \pm 0.03\, M_\odot$ and the values we obtained in this paper including $M_{\mathrm{p}} = 2.27^{+0.49}_{-0.58}\, M_\oplus$, $R_{\mathrm{p}} = 1.37^{+0.10}_{-0.09}\, R_\oplus$, $\emph{P\/} = 0.54929^{+0.00012}_{-0.00005}$ days, and $a_{\mathrm{p}} = 0.0135^{+0.0013}_{-0.0022}$ AU into equation (\ref{equation21}), we can calculate a circularization timescale of $\tau = 8766^{+5556}_{-8116}$ yr, which is a very short time to completely circularize the orbit even considering the large upper uncertainty. Therefore, it is reasonable for us to presume that orbit circularization is achieved for TOI-1807b.

\subsection{Spin-orbit alignment and nodal precession}
\label{section5.4}

   From the parameters of Sector 49 demonstrated in Table~\ref{Table1}, we obtained a transit impact parameter $\emph{b\/} = 0.36^{+0.27}_{-0.25}$. By using the equation $\emph{b\/} = \emph{a\/}\cos\emph{i\/}$, we can calculate the inclination angle $\emph{i\/} = 85.0^{+3.8}_{-3.5}$ deg, which reveals that the orbital plane of TOI-1807b is slightly tilted off the line of sight. Meanwhile, since TOI-1807b is a very young USP planet, there is a great possibility that TOI-1807b is currently undergoing an inclination evolution, where the stellar spin axis and the planetary orbital axis could both be precessed due to their mutual inclination resonances \citep{2019MNRAS.488.3568P}. The nodal precession rate of TOI-1807b driven by the spin axis of TOI-1807 is derived as \citep{2019MNRAS.488.3568P}:
   \begin{equation}
   \begin{aligned}
    {\omega_{\mathrm{nodal}}} = \left(2.7\times10^{-5}\, rad/yr\right)\left(\frac{k_{\mathrm{q}}}{0.01}\right)\\
    \times\left(\frac{a_{\mathrm{p}}}{0.02\, AU}\right)^{-\frac{7}{2}}\left(\frac{M_{\mathrm{\ast}}}{M_{\mathrm{\odot}}}\right)^{-\frac{1}{2}}\left(\frac{R_{\mathrm{\ast}}}{R_{\mathrm{\odot}}}\right)^{5}\left(\frac{P_{\mathrm{\ast}}}{30\, d}\right)^{-2}\,,\label{equation22}
   \end{aligned}
   \end{equation}
   Where $k_{\mathrm{q}}$ is a constant defined by the quadrupole moment of TOI-1807 and is approximately equal to 0.01 for a Sun-like star \citep{2019MNRAS.488.3568P}. If we take $a_{\mathrm{p}} = 0.0135^{+0.0013}_{-0.0022}$ AU, $M_{\mathrm{\ast}} = 0.76 \pm 0.03\, M_\odot$, $R_{\mathrm{\ast}} = 0.690 \pm 0.036\, R_\odot$, and $P_{\mathrm{\ast}} = 8.83 \pm 0.08$ days from Sector 49 parameters, then we can compute the nodal precession rate as $(1.28^{+0.54}_{-0.81}) \times 10^{-2}\, \mathrm{deg\, yr^{-1}}$. This value is about 41 times smaller than the apsidal precession rate of TOI-1807b, indicating that nodal precession is also hardly detectable for a two-year span of observation. Moreover, \citet{2019MNRAS.488.3568P} deduced another equation to calculate the precession rate of the spin axis of TOI-1807 caused by TOI-1807b:
   \begin{equation}
   \begin{aligned}
    {\omega_{\mathrm{spin}}} = \left(7.7\times10^{-7}\, rad/yr\right)\left(\frac{6k_{\mathrm{q}}}{k_{\mathrm{\ast}}}\right)\left(\frac{M_{\mathrm{p}}}{M_{\mathrm{\oplus}}}\right)\\
    \times\left(\frac{a_{\mathrm{p}}}{0.02\, AU}\right)^{-3}\left(\frac{M_{\mathrm{\ast}}}{M_{\mathrm{\odot}}}\right)^{-1}\left(\frac{R_{\mathrm{\ast}}}{R_{\mathrm{\odot}}}\right)^{3}\left(\frac{P_{\mathrm{\ast}}}{30\, d}\right)^{-1}\,,\label{equation23}
   \end{aligned}
   \end{equation}
   Where $k_{\mathrm{\ast}} \sim 0.06$ for a Sun-like star \citep{2019MNRAS.488.3568P} is introduced in equation (\ref{equation17}). By substituting $M_{\mathrm{p}} = 2.27^{+0.49}_{-0.58}\, M_\oplus$, $a_{\mathrm{p}} = 0.0135^{+0.0013}_{-0.0022}$ AU, $M_{\mathrm{\ast}} = 0.76 \pm 0.03\, M_\odot$, $R_{\mathrm{\ast}} = 0.690 \pm 0.036\, R_\odot$, and $P_{\mathrm{\ast}} = 8.83 \pm 0.08$ days into equation (\ref{equation23}), we obtained a value $\omega_{\mathrm{spin}} = (4.81^{+1.90}_{-2.79}) \times 10^{-4}\, \mathrm{deg\, yr^{-1}}$, which is about 27 times smaller than the nodal precession rate of TOI-1807b and $\sim$3 orders of magnitude smaller than the apsidal precession rate of TOI-1807b. Hence, we can safely ignore the precession rate of the spin axis of TOI-1807 and presume that this stellar spin axis always points toward one fixed but unknown direction as we still cannot determine the exact direction of the spin axis of TOI-1807 based on our current data from TESS, which means little is known about whether the stellar spin axis and planetary orbital axis are aligned or misaligned with each other. Nevertheless, the TOI-1807 system is highly possible to realize its spin-orbit alignment since previous trends manifested that hotter stars ($\emph{T\/} > 6200$ K) are more likely to emerge with higher stellar obliquities, resulting in significant spin-orbit misalignment \cite{2010ApJ...718L.145W, 2016ApJ...830....5S}, while TOI-1807 only has a surface temperature of $4757^{+51}_{-50}$ K \citep{Hedges_2021}.

\subsection{Spin-orbit synchronization}
\label{section5.5}

   Spin-orbit synchronization arises from tidal effects. If the planetary orbit is circular (or at least approximately circular), the tidal force would bring the planet to a synchronized period with the star. On the other hand, if the eccentricity of the planet is not negligible, the tidal force will act as a dragging effect at the perihelion, so the rotation of the planet cannot be synchronized with the star. The consequence of being influenced by the drag effect is that the period of the final state will become a value between the undisturbed one and the maximum disturbed one \citep{1965Natur.206R1240P}. The equation of torque caused by the tidal force is \citep{1966Icar....5..375G}:
   \begin{equation}
       N = \frac{9}{4}\left(\frac{GM_{\mathrm{\ast}}^{2}}{Q_{\mathrm{p}}}\right)\left(\frac{R_{\mathrm{p}}^{5}}{a_{\mathrm{p}}^{6}}\right)\,,\label{equation24}
   \end{equation}
   Where \emph{G\/} is the gravitational constant, $Q_{\mathrm{p}}$ is the same tidal quality factor introduced in Section~\ref{section5.3}. From equation (\ref{equation24}), we can see that the maximum drag effect is at the perihelion. We have already mentioned that the orbit of a USP planet is circular. Thus, the torque acting on it will continually change the period of the rotation without any drag effect. Once the rotation has the same period as the orbit, the torque will become 0 (which means the rotational period and orbital period were synchronized), and that is how a USP planet formed its spin-orbit synchronization. In order to evaluate the time for a USP planet to achieve spin-orbit synchronization, \citet{1966Icar....5..375G} introduced another equation about the change of the angular velocity of rotation caused by the tidal force:
   \begin{equation}
       \dot{\omega_{\mathrm{r}}} = -\frac{N}{I} = \frac{45}{8}\left(\frac{GM_{\mathrm{\ast}}^{2}}{M_{\mathrm{p}}Q_{\mathrm{p}}}\right)\left(\frac{R_{\mathrm{p}}^{3}}{a_{\mathrm{p}}^{6}}\right)\,,\label{equation25}
   \end{equation}
   Where $\dot{\omega_{\mathrm{r}}}$ is the change of the angular frequency of the rotation, \emph{N\/} is the torque acting on the planet, $\emph{I\/} = 2M_{\mathrm{p}}R_{\mathrm{p}}^{2}/5$ is the moment of inertia of TOI-1807b, and $Q_{\mathrm{p}}$ is the same tidal quality factor that we introduced in Section~\ref{section5.3} (for the same reasons, we take $Q_{\mathrm{p}}$ = 100 here). Since TOI-1807b orbits in close proximity to its host star, we can expect that TOI-1807b will achieve its spin-orbit synchronization in a very short timescale (where we take $M_{\mathrm{\ast}} = 0.76 \pm 0.03\, M_\odot$, $M_{\mathrm{p}} = 2.27^{+0.49}_{-0.58}\, M_\oplus$, $R_{\mathrm{p}} = 1.37^{+0.10}_{-0.09}\, R_\oplus$, and $a_{\mathrm{p}} = 0.0135^{+0.0013}_{-0.0022}$ AU):
   \begin{equation}
       Timescale \sim \frac{\omega_{\mathrm{r}}}{\dot{\omega_{\mathrm{r}}}} = 0.660^{+0.432}_{-0.688}\, yr\,,\label{equation26}
   \end{equation}
   The large error on $a_{\mathrm{p}}$ places high uncertainties on the timescale, but even so, its upper bound is small enough. Therefore, it is safe to assume that TOI-1807b is tidally locked, which means it is in a synchronous orbit and has a permanent dayside and nightside.

\subsection{Tidal deformation}
\label{section5.6}

   The fact that TOI-1807b is tidally locked leads to permanent deformation of the planet by the elongation of the star-facing axis. The density of TOI-1807b can further be extracted from the radial velocity data of TOI-1807, as well as the transit data we have analyzed to be $0.875^{+0.264}_{-0.285}\, \rho_\oplus$. \citet{2022arXiv220603496N} have given various estimates for the density of TOI-1807b, suggesting a similar composition to Earth. However, these calculations were done assuming a homogeneous spherical model for the planet. Tidal effects, especially for USP planets like TOI-1807, would be very influential both to the orbit and to the planet’s shape and should be considered in its analysis.

   \begin{figure}[ht!]
   \centering
   \includegraphics{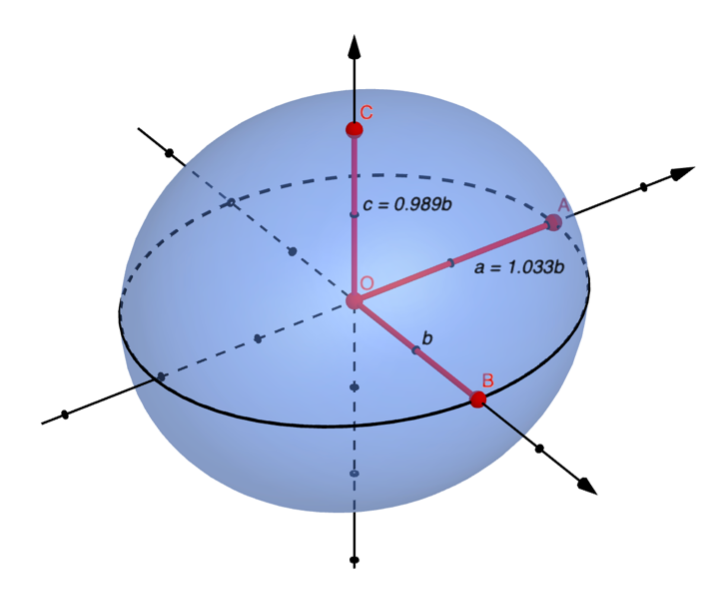}
   \caption{The geometry setup of the distorted TOI-1807b. The longest axis of the ellipsoid points towards the. The c-axis represents the rotational axis.
   \label{Fig11}}
   \end{figure}

   In particular, TOI-1807b is thought to be tidally locked, or spin-orbit synchronized. This allows for permanent deformation of the planet’s shape due to tidal effects in the direction of the star. To construct the planet’s tidal deformation, \citet{2014A&A...570L...5C} developed an analytical model for a planet of uniform density, while \citet{2020ApJ...894....8P} allows variation in inner structure composition. More specifically, due to tides, a tidally locked planet will receive elongation on the axis facing the star. This elongation acts to increase the planet's volume and hence decrease its overall density. Fig.~\ref{Fig11} above demonstrates the geometry of the axis, giving an aspect ratio of 1.04. The method proposed by \citet{2014A&A...570L...5C} makes use of the Love number approach, in which a perturbing potential is proportional to the radial deviation from a sphere. An additional assumption is that the distortion results in an ellipsoid shape. The resulting ellipsoid can be described by a “correction factor” \emph{q\/}:
   \begin{equation}
       q = \frac{k_{\mathrm{p}}}{2}\left(\frac{M_{\mathrm{\ast}}}{M_{\mathrm{p}}}\right)\left(\frac{b}{r_{\mathrm{0}}}\right)^{3}\,,\label{equation27}
   \end{equation}
   Where $r_{\mathrm{0}}$ is the radius of the circular orbit, $k_{\mathrm{p}}$ is the fluid second Love number for radial displacement same as the previous Love number used to calculate the apsidal precession rate of TOI-1807b. The value of the Love number $k_{\mathrm{p}}$ is mainly determined by the inner composition of the planet \citep{2017AJ....154....4P}. With no prior information, however, for rocky planets and Super-Earths (in the case of TOI-1807b), $k_{\mathrm{p}} = 2$ \citep{Yoder95}. The Love number can be more accurately determined if the inner mass composition of the planet is known \citep{Jeffreys_1976}, although for our purposes an estimate is sufficient. The correction factor q is also limited by the Roche limit, and so $\emph{q\/} < k_{\mathrm{p}}/30 = 0.067$. For TOI-1807b specifically, the Roche limit can be calculated as \citep{2014A&A...570L...5C, 2020ApJ...894....8P}:
   \begin{equation}
      r_{\mathrm{R}} = 2.46\left(\frac{M_{\mathrm{\ast}}}{M_{\mathrm{p}}}\right)^{\frac{1}{3}}b = 2.16^{+0.25}_{-0.26}\, R_{\mathrm{\ast}}\,,\label{equation28}
   \end{equation}
   The Roche limit is at 2.16 stellar radii, and so (as expected) TOI-1807b will avoid tidal disruption because TOI-1807b is safely outside the Roche limit. The tidal distortions, however, can still be described as an ellipsoid:
   \begin{eqnarray}
      a = \left(1 + 3q\right)b\,,\label{equation29}\\
      c = \left(1 - q\right)b\,,\label{equation30}
   \end{eqnarray}
   Where $\emph{b\/} > \emph{c\/}$ because of the centrifugal potential. We performed this analysis on TOI-1807b. TOI-1807b, from our calculations, has a mass of 2.27 Earth masses, while our analysis on Sector 49 computes $\emph{a\/} = a_{\mathrm{p}}/R_{\mathrm{\ast}} = 4.20$ and $\emph{k\/} = R_{\mathrm{p}}/R_{\mathrm{\ast}} = 0.0183$ and suggests that TOI-1807b’s shadow would be largely spherical during transit so that $R_{\mathrm{p}} \approx \emph{b\/}$ and \emph{q\/} = 0.01. To first order, the density correction is given by \citet{2014A&A...570L...5C} as:
   \begin{equation}
      \frac{\rho}{\rho_{\mathrm{s}}} = \left(1 - 3.5q + 6q\cos^2{i}\right)\,,\label{equation31}
   \end{equation}
   Where $\rho_{\mathrm{s}}$ is the density calculated assuming a spherical model and \emph{i\/} is the inclination of the system. TOI-1807b can be assumed to be viewed edge-on \citep{2022arXiv220603496N}, $\rho = 0.965\, \rho_{\mathrm{s}}$. This results in $\sim$4\% correction to TOI-1807b’s density. This correction may very well be absorbed into parameter errors. Fitted values for Sector 22, 23, and 49 give an uncertainty of 3--4\% in the value of the planetary radius, and any additional uncertainties introduced will absorb the 4\% correction in density. However, if higher resolution light curves can be obtained, further constraints can be placed on the size of TOI-1807b to deduce its true density. Additional simulations such as those by \citet{2020ApJ...894....8P} can further lead us to deduce the internal structure of lava worlds and their similarities to Earth.

\subsection{Relativistic effect}
\label{section5.7}

\begin{figure}[ht!]
   \centering
   \includegraphics{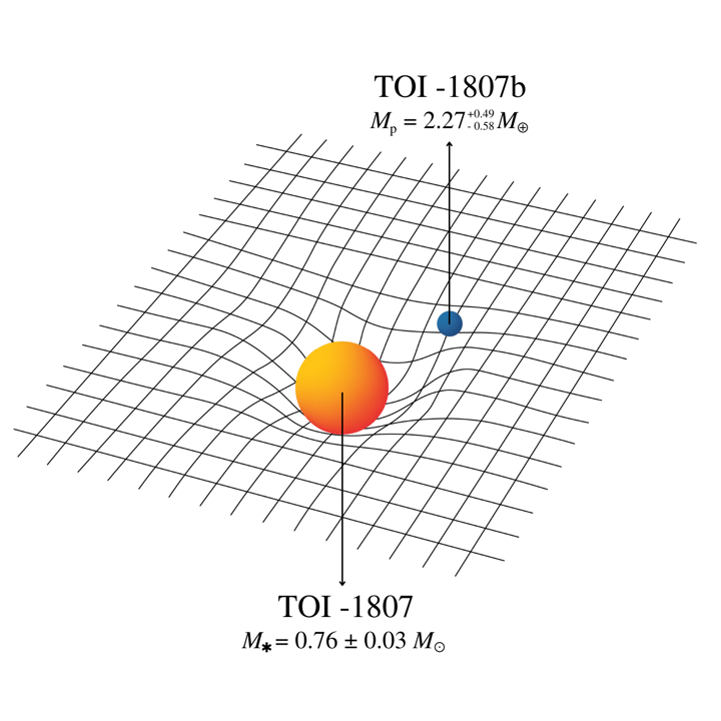}
   \caption{The visualization of the GR effect. The gravitational potential of TOI-1807 will change its surrounding space, and the direct result of this is the gravitational redshift. The change of the precession angle $\varphi$ will be a more complicated case, which is derived from the GR equation in a circular motion.
   \label{Fig12}}
   \end{figure}

   In this section, we discuss the General Relativistic (GR) apsidal precession of TOI-1807b. The special relativistic effect can be ignored since the orbital speed of TOI-1807b is much smaller than the speed of light in vacuum space. Prior to special relativity, the old notion of space and time was overthrown in GR. The massive body will change its surrounding space due to the equivalence principle as illustrated in Fig.~\ref{Fig12}, which is expressed as the statement that the gravitational force is proportional to the inertial mass in Newtonian theory. A direct consequence of the equivalence principle is just the gravitational redshift.

   In our Solar System, the apsidal precession of Mercury is famously known to confirm the GR theories, which perfectly explain the discrepancy of $\sim$43" per century between the theoretical prediction and actual observation. Corresponding to the description that we had in the case of Mercury, TOI-1807b is a rocky planet with an orbital distance of around 0.0135 AU (much smaller than Mercury’s orbital distance of 0.387 AU), and the mass of TOI-1807 is $0.76 \pm 0.03\, M_\odot$ (which is great enough to produce a strong GR effect). The preceding discussion implies that TOI-1807b may have a stronger relativistic effect than planet Mercury, and we might detect some GR apsidal precession of TOI-1807b by measuring a time variation in the epoch of transits and occultations \citep{2021MNRAS.505.1567A}.

   The prograde precession of the argument of periastron given by GR is \citep{2021MNRAS.505.1567A}:
   \begin{equation}
       \delta\varphi = \frac{6\pi}{a_{\mathrm{p}}\left(1 - e^2\right)}\frac{GM}{c^2}\,,\label{equation32}
   \end{equation}
   Where \emph{M\/} represents the mass of the total planetary system (which is equal to the sum of the mass of the star and the mass of the planet), \emph{e\/} is the orbital eccentricity, and \emph{c\/} is the speed of light in vacuum space. Once we substitute the mass of the TOI-1807 system, and the orbital eccentricity \emph{e\/} of TOI-1807 into the above equation, we can calculate the perihelion precession rate of TOI-1807b. The equation to compare the result of TOI-1807b to the semi-major axis GR apsidal precession of Mercury is written below \citep{2022arXiv220603496N}:
   \begin{equation}
       \delta\varphi = \delta\varphi_{\mathrm{Mer}}\frac{a_{\mathrm{Mer}}\left(1-e_{\mathrm{Mer}}^{2}\right)}{a_{\mathrm{p}}}\left(\frac{M_{\mathrm{\ast}}+M_{\mathrm{p}}}{M_{\mathrm{\odot}}+M_{\mathrm{Mer}}}\right)\,,\label{equation33}
   \end{equation}
   Where $M_{\mathrm{\ast}}$ is the mass of TOI-1807, $M_{\mathrm{p}}$ is the mass of TOI-1807b calculated from Sector 49 data (where we take $M_{\mathrm{\ast}} = 0.76 \pm 0.03\, M_\odot$, $M_{\mathrm{p}} = 2.27^{+0.49}_{-0.58}\, M_\oplus$, $a_{\mathrm{p}} = 0.0135^{+0.0013}_{-0.0022}$ AU). Since the orbit of TOI-1807b is regarded to be circular (\emph{e\/} = 0), the eccentricity term of the denominator in equation (\ref{equation33}) is then removed. Based on the two equations above, we can compute a result of $\sim$2.37 arcsec per period for the perihelion precession rate of TOI-1807b, which gives a 23 times larger value than the perihelion precession rate measured for Mercury ($\sim$0.104 arcsec per period). The GR apsidal precession of TOI-1807b we obtained in this paper corrects the previous value of 0.23 arcsec per period published by \citet{2022arXiv220603496N}.

   Another way to detect relativistic effects is to measure the gravitational redshift. Gravitational redshift was observed by the Pound-Rebka-Snider experiment in 1959 for the first time \citep{2018PhLA..382.2192L}. The idea is to check whether the frequency of the electromagnetic wave was influenced by the gravitational potential by setting an atomic clock on a rocket-launched satellite. The difference in the frequency $\Delta\nu$ between two atomic clocks is defined by the formula \cite{2006LRR.....9....3W}:
   \begin{equation}
       z = \frac{\Delta\nu}{\nu} = \left(1+\alpha\right)\frac{\Delta U}{c^2}\,,\label{equation34}
   \end{equation}
   Where $\nu$ is the transition frequency of the atomic clock on the ground, $\Delta U$ is the difference of gravitational potential between two points where one point is on the ground and another point is on the satellite, and $\alpha$ is the parameter introduced to testify GR (when $\alpha$ = 0, GR is valid). In this formula, we see that if the difference in gravitational potential between two points is huge, then it is highly possible to detect the gravitational redshift. From Newtonian Mechanics, we know that $\varphi_{p} = -GM/r$, and considering that the mass of Earth is much smaller than the mass of TOI-1807, we can ignore the gravitational potential of the Earth. From the parameters we obtained in Sector 49 ($M_{\mathrm{\ast}} = 0.76 \pm 0.03\, M_\odot$, $a_{\mathrm{p}} = 0.0135^{+0.0013}_{-0.0022}$ AU), we can calculate the gravitational redshift of TOI-1807b as $\emph{z\/} = (5.53^{+0.06}_{-0.09}) \times 10^{-7}$, which is very hard to detect. It is hence unlikely that the transit signals are distorted by the gravitational redshift.

\subsection{Formation theories}
\label{section5.8}

   Currently, our main planetary formation theories stem from circumstellar disks, which surround a newly formed star. This disk is composed of mainly hydrogen and helium gas, with only a small percentage ($\sim$~1\%) that is solid. This material undergoes accretion and slowly forms into planets. Younger star systems are especially useful for this purpose as they are more likely to retain properties from the protoplanetary disks around a new star. TOI 1807b, as the youngest USP planet discovered up to date, allows us to study the properties of a Super-Earth planet that has perhaps recently lost its atmosphere \citep{Hedges_2021}. Earlier in the paper, we had covered the circularization caused by tidal dissipation and it is a potential reason for the formation of USP planets \citep{2010ApJ...724L..53S, 2017ApJ...842...40L, 2020ApJ...905...71M}, as well as USP planets being the result of inward migration within the planetary system, there exists a range of other possible explanations. \citet{2008MNRAS.384..663R} discussed five other possible ways of formation for ‘hot Earths’ as follows:
   \begin{enumerate}
       \item In situ formation: It may be possible that accretion occurs near the central star from protoplanetary disks where the mass concentration near the center ($< 1$ AU) is higher.
       \item Shepherding from gas giant migration: Hot Earths formed because of the inward migration of gas giant planets, which pulls material inwards. This would result in USP planets next to gas giants. However, no transiting gas giants are found from the light curve of TOI-1807, which puts doubt on this theory. Even if out-of-transit gas giants are present, their gravitational effects on the period of TOI-1807b should be observable.
       \item Shepherding from sweeping secular resonance: This results in the coexistence of the hot Earth with at least two other giant planets. As above, no other planets were found from the transit curve, although further analysis of the RV curves of TOI-1807 can allow better confidence.
       \item Migration of a gas planet and photo-evaporation: A planet that was formerly a gas planet has migrated inwards, perhaps too close to the star. This migration can cause them to lose a portion of their gas surface due to irradiative XUV heating \citep{Lammer_2003, 2004A&A...419L..13B}.
       \item Inward migration of Earth or Super-Earth sized planet within a planetary system: With the commonality of multi-planetary systems, there exists the potential for more discoveries of planets orbiting TOI-1807, thus providing a better opportunity to study the effects of other planets on the formation of USP planets.
   \end{enumerate}
   
\section{Conclusions}
\label{section6}

   In this paper, we reviewed the “lava world” TOI-1807b by analyzing its transit data from three TESS sectors: Sector 22, 23, and 49. We confirmed the presence of a transiting planet by developing our own flattening mechanism to remove stellar variations in all three sectors, and we deduced the transit parameters by using an MCMC algorithm to fit the latest Sector 49 data. From these parameters, we explored various properties of this lava world. As a result, the parameters agreed reasonably well across all three sectors, without any significant deviations. We also used our results to calculate the orbital decay rates and confirmed the tidal circularization of TOI-1807b, suggesting that the planet is tidally locked as thought and is unlikely to undergo orbital decay at present. This raises the possibility of the planet being tidally deformed because of spin-orbit synchronization, and we further investigated the potential corrections of the planet's radius and density induced by tidal deformations. However, we did find our results to be somewhat inconsistent with those by \citet{2022arXiv220603496N}, and our parameters are not as well constrained. Without the use of RV data, the transit method to derive the planetary parameters of TOI-1807b will yield a poorer result. To be more specific, we note that the {\fontfamily{pcr}\selectfont mandelagol} transit model cannot put a tight constraint on the impact parameter b, which is probably due to the short duration of the transits. We were unable to deduce the stellar limb darkening coefficients as well, which might be further ameliorated by using non-linear law to better determine the stellar limb darkening coefficients. To better constrain the parameters and calculate the stellar limb darkening coefficient, another Python package called {\fontfamily{pcr}\selectfont batman} is suggested to implement to model the transit light curves more precisely \citep{2015PASP..127.1161K}.

   There are many remarkable prospects to explore in the future. Firstly, to extrapolate the composition of TOI-1807 more rigorously, a simulation developed by \cite{1986aApJS...61..479H, 1986bApJS...62..461H} could be used to quantitatively estimate the iron core mass fraction and silicate mantle mass fraction \citep{2020ApJ...894....8P}. In addition, considering that it is unlikely for a single USP planet to exist in a planetary system without any other outer companions, further research with more advanced observation equipment such as the James Webb Space Telescope (JWST) might be required to obtain and analyze more accurate data so that we can possibly discover a second exoplanet in the TOI-1807 system, and the USP planets formation theory can also be improved with more details or even be confirmed in the future. Although in Section~\ref{section5.4} we failed to determine the spin-orbit alignment in the TOI-1807 system, we are still able to indirectly ascertain it by analyzing the Rossiter-McLaughlin (RM) effect arising from stellar rotation, which is an alternative but effective method to detect the transiting planets and to confirm a prograde or a retrograde orbit \citep{2010arXiv1001.2010W}. What’s an even more intriguing feature we didn't cover in this paper is the possible planetary obliquity of TOI-1807b when the planetary spin axis is tilted off the orbital axis, which is another tidal dissipation source to accelerate the tidal orbital decay \citep{2020ApJ...905...71M} and hence may further impact the future evolution of TOI-1807b.


\section{Acknowledgements}
This paper includes data collected by the TESS mission. Funding for the TESS mission is provided by NASA's Science Mission Directorate. We acknowledge the use of public TESS Alert data from pipelines at the TESS Science Office and at the TESS Science Processing Operations Center. This research has made use of the NASA Exoplanet Archive, which is operated by the California Institute of Technology, under contract with the National Aeronautics and Space Administration under the Exoplanet Exploration Program. This research made use of {\fontfamily{pcr}\selectfont Lightkurve}, a Python package for Kepler and TESS data analysis, and its dependencies, which include {\fontfamily{pcr}\selectfont Astropy}, {\fontfamily{pcr}\selectfont Astroquery}, and {\fontfamily{pcr}\selectfont Tesscut}. We also wish to thank Professor Joshua N. Winn for providing useful suggestions and support for our research. Thank you to Tuochen Gong for your assistance with our queries and your support during our research.

\software{Numpy \citep{harris2020array},
          Matplotlib \citep{Hunter:2007},
          Scipy \citep{2020SciPy-NMeth},
          Lightkurve \citep{2018ascl.soft12013L},
          Astropy \citep{2018AJ....156..123A},  
          Astroquery \citep{Ginsburg_2019}, 
          Tesscut \citep{2019ascl.soft05007B}
          }

\bibliography{Final_Paper}
\bibliographystyle{aasjournal}       



\begin{appendix}
\section{Transit visualization in the light curves of Sector 22 and 23}
\label{AppendixA}

\begin{figure*}[ht!]
   \centering
   \includegraphics[width=17.3cm]{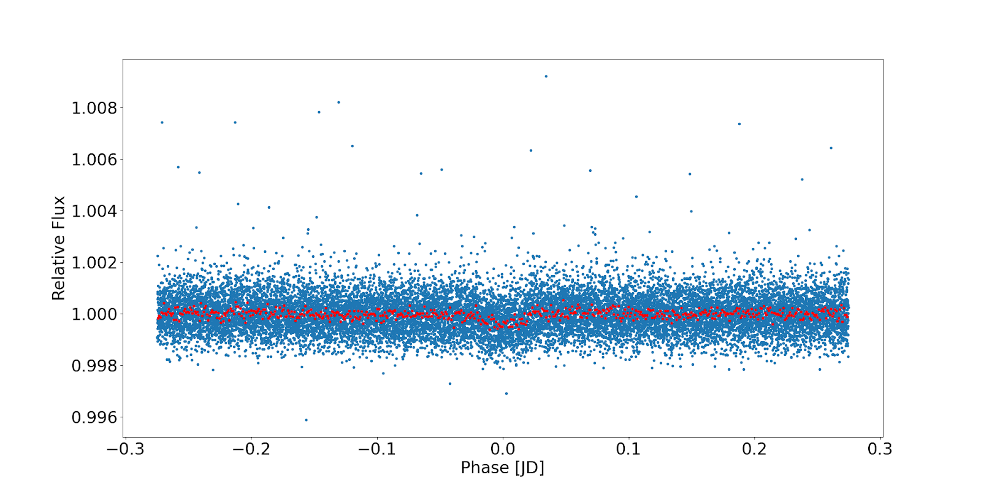}
   \caption{The flattened, phase-folded light curves of Sector 22 which is plotted for transit visualization.}
   \label{FigA1}
   \includegraphics[width=17.3cm]{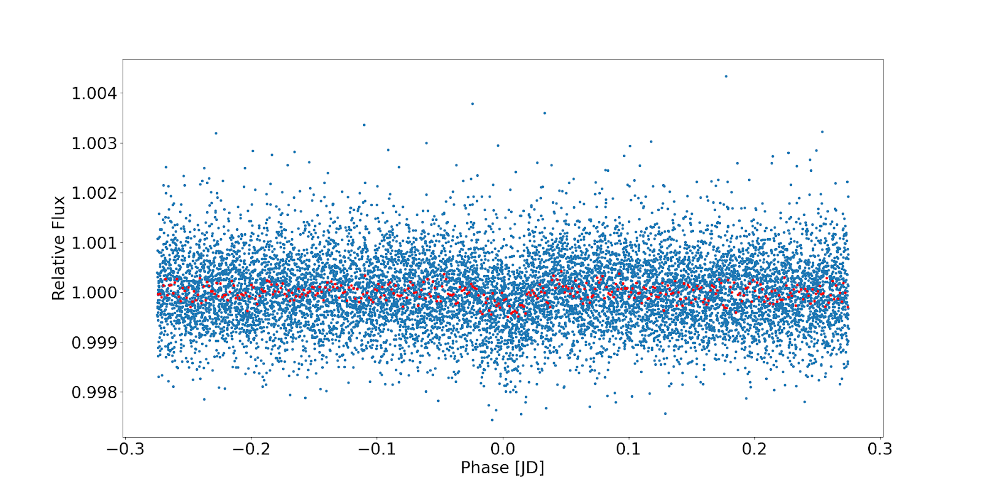}
   \caption{The flattened, phase-folded light curves of Sector 23 which is plotted for transit visualization.}
   \label{FigA2}
   \end{figure*}

   In Fig.~\ref{FigA1} and Fig.~\ref{FigA2}, we plotted the transit visualization for the light curves of Sector 22 and 23 respectively, where each data point in red represents the averaged position of 30 data points in blue.

\section{The fits for the light curves of Sector 22 and 23}
\label{AppendixB}

\begin{figure*}[ht!]
   \centering
   \includegraphics[width=17.3cm]{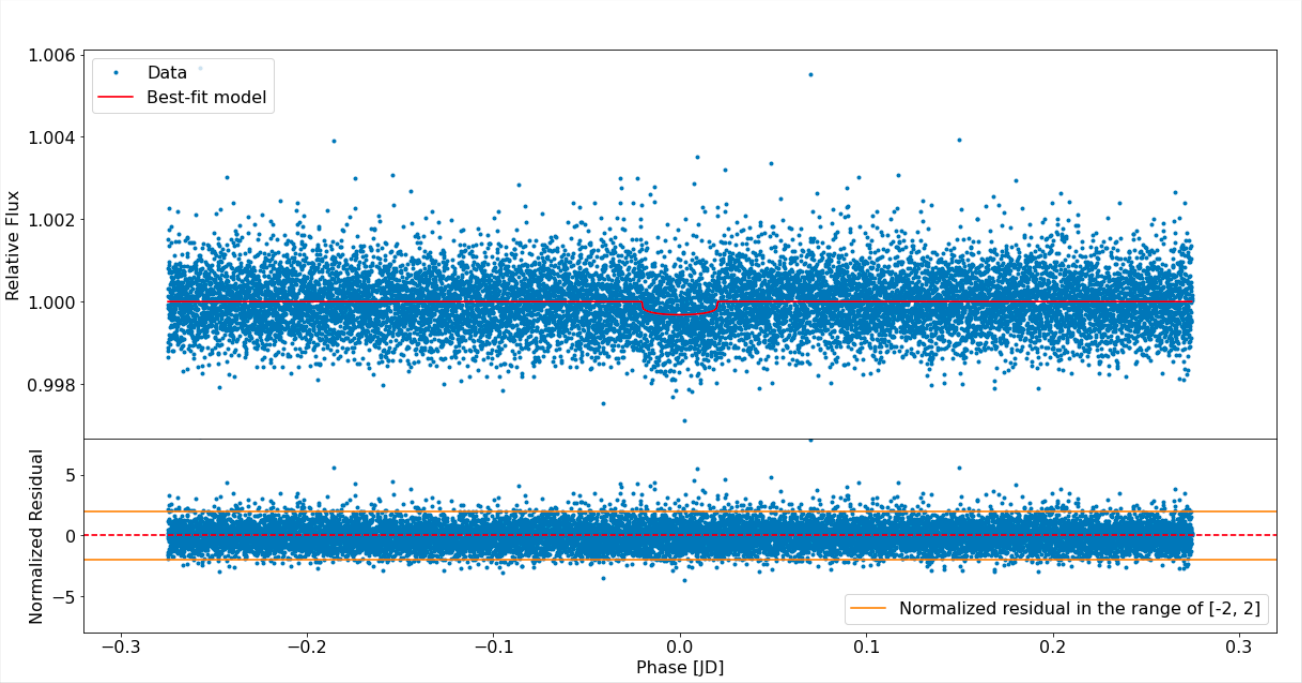}
   \caption{The light curve fitted by the transit model as well as the normalized residual plot for Sector 22.}
   \label{FigB1}
   \includegraphics[width=17.3cm]{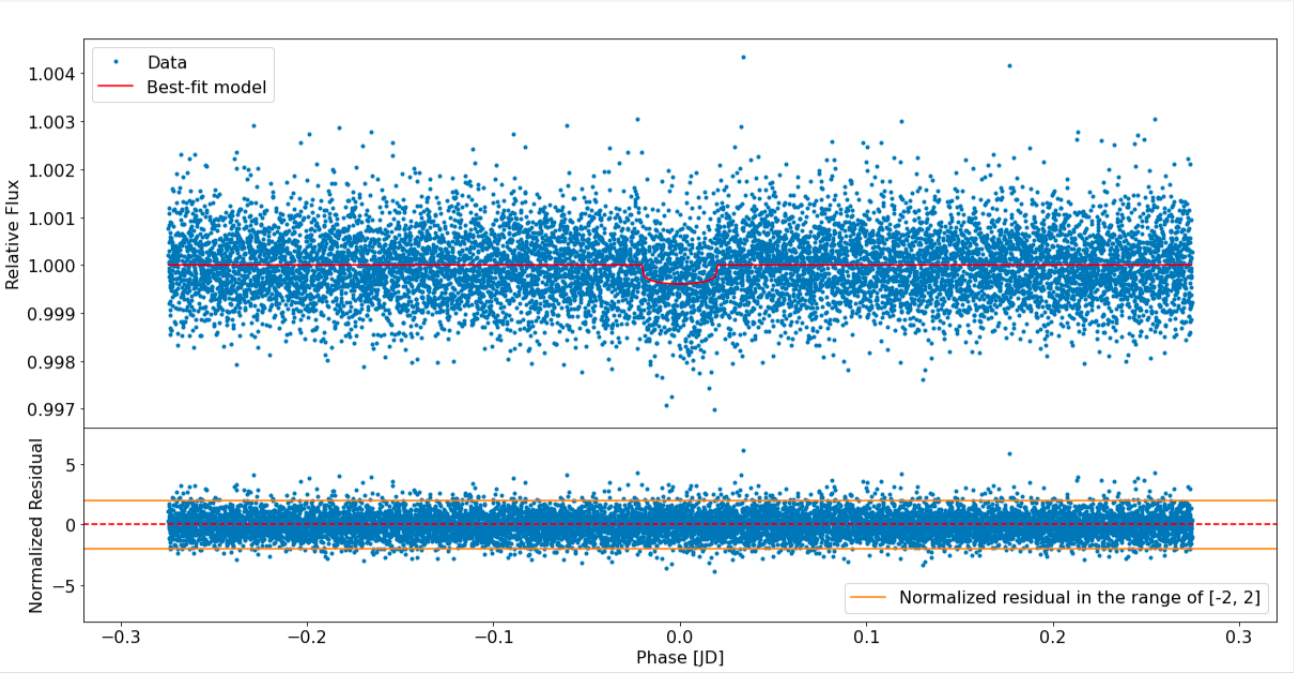}
   \caption{The light curve fitted by the transit model as well as the normalized residual plot for Sector 23.}
   \label{FigB2}
   \end{figure*}

   In Fig.~\ref{FigB1} and Fig.~\ref{FigB2}, we plotted the transit model fits for both the light curves of Sector 22 and 23. The normalized residual plot to each light curve plot was included underneath, where the central red dashed line represents the normalized residual with the value of 0. One should notice that 96\% data points are contained within $\pm 2$, which displays a good fit.

\section{The parameters of Sector 22 and 23 as well as their corner plots}
\label{AppendixC}

\begin{figure*}[ht!]
   \centering
   \includegraphics[width=17.3cm]{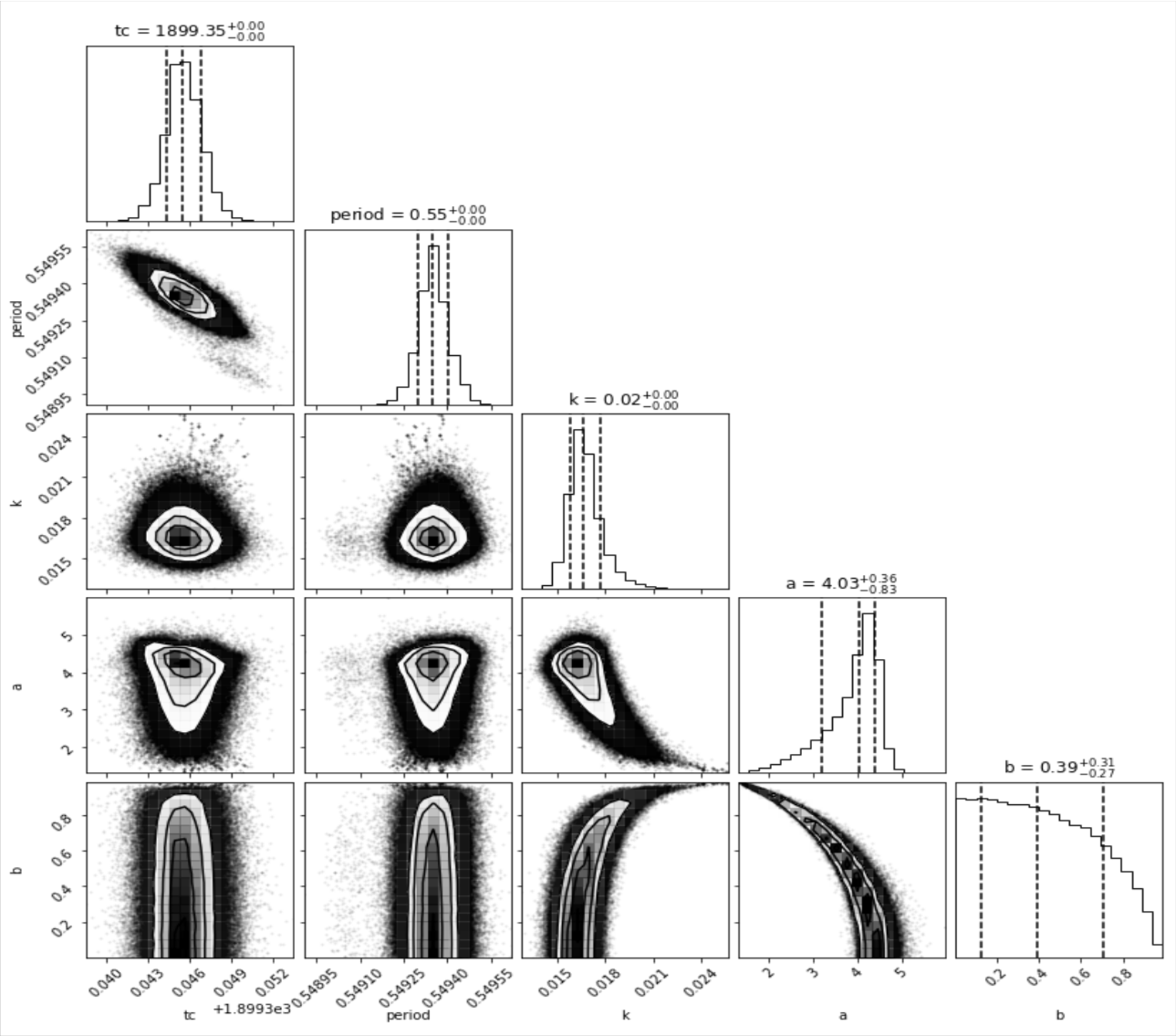}
   \caption{The corner plot of Sector 22 parameters obtained from the MCMC algorithm.}
   \label{FigC1}
   \end{figure*}

\begin{figure*}[ht!]
   \centering
   \includegraphics[width=17.3cm]{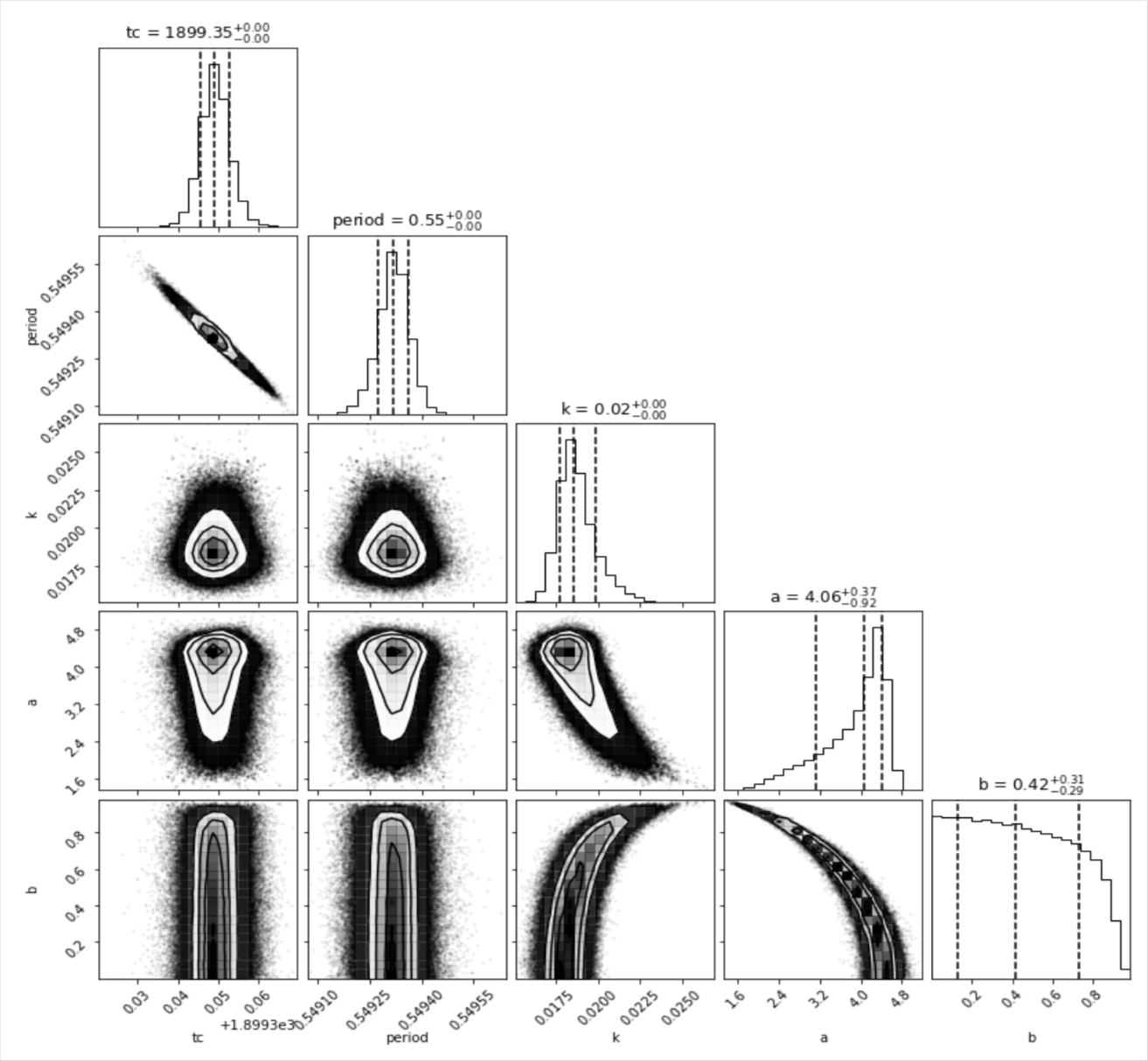}
   \caption{The corner plot of Sector 23 parameters obtained from the MCMC algorithm.}
   \label{FigC2}
   \end{figure*}

   In Table~\ref{TableC1} and Table~\ref{TableC2}, we further recorded the parameters of Sector 22 and 23 calculated from both the curve fit method and the MCMC algorithm, while Fig.~\ref{FigC1} and Fig.~\ref{FigC2} illustrated the corner plots of Sector 22 and 23 parameters after running through the MCMC algorithm.

   \renewcommand{\arraystretch}{1.2}
   \begin{table*}[ht!]
        \begin{tabular}{p{7cm} p{1cm} p{3cm} p{3cm}}
        \hline\hline
        Parameter & Unit & MCMC result & Optimal value \\
        \hline
            Conjunction time ($t_{\mathrm{c}}$) & BTJD & $1899.34547^{+0.00131}_{-0.00116}$ & 1899.34493   \\
            Period (\emph{P\/}) & days & $0.54935 \pm 0.00005$ & 0.54935 \\
            Planetary radius to stellar radius ratio (\emph{k\/}) & / & $0.01660^{+0.00108}_{-0.00081}$ & 0.0162 \\
            Orbital semi-major axis to stellar radius ratio (\emph{a\/}) & / & $4.03^{+0.35}_{-0.83}$ & 4.39  \\
            Transit impact parameter (\emph{b\/}) & / & $0.39^{+0.31}_{-0.27}$ & / \\
        \hline
        \end{tabular}
        \caption{The parameters of Sector 22 obtained from both the curve fit method and the MCMC algorithm.}
        \label{TableC1}
   \end{table*}

   \begin{table*}[ht!]
        \begin{tabular}{p{7cm} p{1cm} p{3cm} p{3cm}}
        \hline\hline
        Parameter & Unit & MCMC result & Optimal value \\
        \hline
            Conjunction time ($t_{\mathrm{c}}$) & BTJD & $1899.34912^{+0.00357}_{-0.00339}$ & 1899.35014   \\
            Period (\emph{P\/}) & days & $0.54935^{+0.00004}_{-0.00005}$ & 0.54930 \\
            Planetary radius to stellar radius ratio (\emph{k\/}) & / & $0.01860^{+0.00124}_{-0.00086}$ & 0.0189 \\
            Orbital semi-major axis to stellar radius ratio (\emph{a\/}) & / & $4.06^{+0.37}_{-0.92}$ & 3.80  \\
            Transit impact parameter (\emph{b\/}) & / & $0.42^{+0.31}_{-0.29}$ & / \\
        \hline
        \end{tabular}
        \caption{The parameters of Sector 23 obtained from both the curve fit method and the MCMC algorithm.}
      \label{TableC2}
   \end{table*}

\end{appendix}




\end{document}